\documentclass[12pt]{article}
\usepackage{graphicx}
\usepackage{latexsym}
\usepackage{ifpdf}
\ifpdf
   \usepackage{textcomp}
\usepackage{amssymb}
\usepackage{amsfonts,amsthm,amstext,amscd}
\usepackage{slashed}
\usepackage{amsmath}

\newcommand{\beqa}{\begin{eqnarray}}
\newcommand{\eeqa}{\end{eqnarray}}

\newcommand{\be}{\begin{equation}}
\newcommand{\ee}{\end{equation}}
\newcommand{\beq}{\begin{equation}}
\newcommand{\eeq}{\end{equation}}
\newcommand{\bea}{\begin{eqnarray}}
\newcommand{\eea}{\end{eqnarray}}

\newcommand{\bear}{\begin{eqnarray}}
\newcommand{\eear}{\end{eqnarray}}
\begin{document}
\baselineskip=15.5pt
\pagestyle{plain}
\setcounter{page}{1}

\def\r{\rho}
\def\CC{{\mathchoice
{\rm C\mkern-8mu\vrule height1.45ex depth-.05ex
width.05em\mkern9mu\kern-.05em}
{\rm C\mkern-8mu\vrule height1.45ex depth-.05ex
width.05em\mkern9mu\kern-.05em}
{\rm C\mkern-8mu\vrule height1ex depth-.07ex
width.035em\mkern9mu\kern-.035em}
{\rm C\mkern-8mu\vrule height.65ex depth-.1ex
width.025em\mkern8mu\kern-.025em}}}

\newfont{\namefont}{cmr10}
\newfont{\addfont}{cmti7 scaled 1440}
\newfont{\boldmathfont}{cmbx10}
\newfont{\headfontb}{cmbx10 scaled 1728}
\renewcommand{\theequation}{{\rm\thesection.\arabic{equation}}}
\font\cmss=cmss10 \font\cmsss=cmss10 at 7pt
\par\hfill DFTT 32/2011

\begin{center}
{\LARGE{\bf Unbalanced Holographic Superconductors \\and Spintronics}}
\end{center}
\vskip 10pt \vskip 10pt
\begin{center}
{\large
Francesco Bigazzi $^{a,b}$, Aldo L. Cotrone $^{c}$, Daniele Musso $^{c}$, \\
Natalia Pinzani Fokeeva$^{a,d,e}$, Domenico Seminara $^{a,d}$.}
\end{center}
\vskip 10pt
\begin{center}
\textit{$^a$ Dipartimento di Fisica e Astronomia, Universit\`a di
Firenze; Via G. Sansone 1, I-50019 Sesto Fiorentino
(Firenze), Italy.}\\
\textit{$^b$ INFN, Sezione di Pisa; Largo B. Pontecorvo, 3, I-56127
Pisa, Italy.}\\
\textit{$^c$  Dipartimento di Fisica, Universit\'a di Torino and
INFN Sezione di Torino; \\Via P. Giuria 1, I-10125 Torino, Italy.}\\
\textit{$^d$ INFN, Sezione di Firenze; Via G. Sansone 1, I-50019
Sesto Fiorentino (Firenze), Italy. }\\
\textit{$^e$ Institute for Theoretical Physics, University of Amsterdam;
Science Park 904, Postbus 94485, 1090 GL Amsterdam, The Netherlands.}\\
\vspace{7pt}
{\small bigazzi@fi.infn.it, cotrone@to.infn.it, mussod@to.infn.it,
npinzani@fi.infn.it, seminara@fi.infn.it}
\end{center}
\vspace{15pt}

\begin{center}
\textbf{Abstract}
\end{center}

\vspace{4pt}{\small \noindent We present a minimal holographic model for s-wave superconductivity with
unbalanced Fermi mixtures, in $2+1$ dimensions at strong coupling.  The breaking of a $U(1)_A$
``charge'' symmetry is driven by a non-trivial profile for a charged scalar field in a charged
asymptotically $AdS_4$ black hole. The chemical potential imbalance is implemented by turning on the
temporal component of a $U(1)_B$ ``spin'' field under which the scalar field is uncharged. We study
the phase diagram of the model and comment on the eventual (non) occurrence of LOFF-like
inhomogeneous superconducting phases. Moreover, we study ``charge'' and ``spin'' transport, implementing
a holographic realization (and a generalization thereof to superconducting setups) of Mott's two-current model which provides the theoretical basis of modern spintronics. Finally we comment on
possible string or M-theory embeddings of our model and its higher dimensional generalizations, within
consistent Kaluza-Klein truncations and brane-anti brane setups.} \vfill
\newpage


\section{Introduction and overview}
\setcounter{equation}{0} The occurrence of superconductive phases where two fermionic species
contribute with unbalanced populations, or unbalanced chemical potentials, is an interesting
possibility relevant both in condensed matter and in finite density QCD contexts (see \cite{casnar}
for a review). A chemical potential mismatch between the quarks is naturally implemented in high
density QCD setups due to mass and charge differences between the quark species. In metallic
superconductors the imbalance can be induced by e.g. paramagnetic impurities, modeled by means of a
Zeeman coupling of the spins of the conducting electrons with an effective external magnetic
``exchange'' field.

At weak coupling, unbalanced Fermi mixtures are expected to develop
 inhomogeneous superconducting phases, where the Cooper pairs
have non-zero total momentum. This is the case of the Larkin-Ovchinnikov-Fulde-Ferrel (LOFF) phase
\cite{loff}. The latter can develop, at small values of the temperature, provided the chemical
potential mismatch $\delta\mu$ is not too large (if the Fermi surfaces of the two species are too far
apart, the system reverts to the normal non-superconducting phase) and not below a limiting value
$\delta\mu=\delta\mu_{1}$ found by Chandrasekhar and Clogston \cite{cc}. At this point, at zero
temperature, the system experiences a first order phase transition between the standard
superconducting and the LOFF phase.
The experimental occurrence of such inhomogeneous phases is still
unclear, and establishing their appearance in strongly-coupled
unconventional systems from a theoretical point of view is a challenging
problem.

The imbalance of spin populations in ferromagnetic materials is also a relevant ingredient of
spintronics, the branch of electronics concerned with the study and applications of spin transport.
Spintronics constitutes a very important research area, due to its technological outcomes, e.g. in computer hard disk devices. From
the theoretical point of view, the research in this area is vast and quickly developing. See
\cite{nobel,review} and references therein for interesting introductions to the subject.

The roots of spintronics are based on Mott's ``two-current model'' \cite{mott}. Mott's model describes
ferromagnetic materials at low temperature and in its simplest version it treats both charge and spin
as conserved quantities, neglecting dissipative spin-flip interactions like spin-orbit terms. In the
model, charge and spin currents flow in parallel and can be both induced either by turning on an
electric field or a ``spin motive force'' \cite{fert1,vanson}. The latter can be practically realized
by a space gradient in the chemical potential imbalance, for example by means of a ferromagnetic-non
ferromagnetic junction. The electric and spin motive forces can be described by means of two
(``effective'' in the case of the spin\footnote{In $3+1$ dimensions, the $U(1)$ spin symmetry is
contained in the $SU(2)$ spin group which is a good approximate symmetry at low energy and
temperature, see e.g. \cite{vanson,iqbal}.}) $U(1)$ gauge fields. This picture has been recently used
in the system of degenerate free electrons subjected to impurity scattering in ferromagnetic conductors
\cite{shibataealtro}. It is an interesting problem to see how the two-current model can be realized in
general setups (even superconducting ones) where a weakly coupled quasi-particle description cannot be
employed.

With the aim of providing some toy-model-based insights on these issues, we have studied the simplest
holographic realization of strongly coupled unbalanced s-wave superconductors in the grand-canonical
ensemble at non-zero temperature.\footnote{For an introduction to the literature on holographic
studies of condensed matter systems see the reviews \cite{reviewsADSCM} and references therein.}
Experimental evidence suggests that high $T_c$ superconductors like certain cuprates are effectively
layered. Moreover it is expected that they develop a quantum critical point (QCP) in their phase
diagram \cite{sachdev}. If the QCP displays conformal invariance, the physics at this point (and
within the so called critical region) is effectively described by a $2+1$ dimensional conformal field
theory at strong coupling (and at finite temperature and density). A related holographic description,
within a simple bottom-up approach, is provided by a gravitational dual model in $3+1$ dimensions with
the following minimal ingredients. The breaking of a $U(1)_A$ ``charge'' symmetry characterizing
superconductivity is driven, on the gravity side, by a non-trivial scalar field charged under a
$U(1)_A$ Maxwell field in an asymptotically $AdS_4$ black hole background as in
\cite{gubser,hhh1,hhh2}. The chemical potential mismatch, which can also be interpreted as a chemical
potential for a $U(1)_B$ ``spin'' symmetry (see also \cite{iqbal}), is accounted for in the gravity setup by turning on the
temporal component of a Maxwell field $U(1)_B$ under which the scalar field is
uncharged.\footnote{Similarly, the holographic unbalanced p-wave setup of \cite{erdmeng} contains
$SU(2)\times U(1)_B$ gauge fields. An $U(1)_3\subset SU(2)$ symmetry is broken by the condensation of e.g.
a $U(1)_1\subset SU(2)$ vector \cite{gubserpwave} which is uncharged under $U(1)_B$.}

The model depends on two parameters, namely the charge $q$ of the scalar field and its mass $m$. For a
particular choice of the latter (namely $m^2=-2$ in units where the $AdS$ radius is set to one), aimed
at implementing a fermionic condensate of canonical dimension $2$,\footnote{In $2+1$ dimensions a weakly coupled Cooper pair has dimension 2.} we will show that the critical
temperature below which a superconducting homogeneous phase develops decreases with the chemical
potential mismatch, as it is expected in weakly coupled setups. However, the phase diagram arising
from the holographic model shows many differences with respect to its weakly coupled counterparts. In
particular there is no sign of a Chandrasekhar-Clogston bound at zero temperature and the
superconducting-to-normal phase transition is always second order. This leads us to argue that a LOFF
phase should not show up (we have checked that this is actually the case in the $q\gg1$ limit). A
different situation arises for different choices of the parameters which seem to allow for
Chandrasekhar-Clogston-like bounds; we will not explore these choices in detail in this paper.

Using standard holographic techniques, we also study ``charge'' and ``spin'' transport in our model.
Essentially we turn on an external electric field $E^A$ as well as a ``spin motive field'' $E^B$ and
look at the correspondingly generated ``charge'' and ``spin'' currents $J^A$ and $J^B$. The
holographic model provides a quite natural realization of Mott's two-current model.\footnote{Different
holographic models with two currents have been studied in the literature. Nevertheless, as far as we
are aware of, the connection with the two-current model has not been explored so far.}
It also automatically provides a non-trivial conductivity matrix for the optical ``charge'', ``spin''
and other conductivities (we study heat transport, too) at zero momentum, both in the superconducting and in the normal phase (which
can be also seen as a holographic ``forced'' ferromagnetic phase \cite{iqbal}). The intertwining of spin and
charge transport is mediated, holographically, by the interaction of the $U(1)_A$ and $U(1)_B$ Maxwell
fields with the metric.\footnote{In our model this is the only way in which the two Maxwell fields can
couple. In the dual field theory this implies that the corresponding currents are only coupled through
``gluon-like'' loops. This feature is model dependent: in other holographic setups, such as those with
two overlapping ``flavor'' D-branes, the coupling can happen also more directly, e.g. by means of non linear
terms in the brane action \cite{erdmeng} or by $F_A\wedge F_B$ terms \cite{stripped}. The coupling via the energy-momentum tensor operator is nevertheless generically present in these cases too.} This very
general simple phenomenon, not related with particular assumptions in our model, suggests that
analogous ``spintronics'' effects could occur in generic strongly coupled unbalanced models, for
example in QCD at finite baryon and isospin density.

The phenomenological bottom-up holographic approach we adopt in this paper, aims at
providing information on some universal properties of classes of strongly coupled field theories with
the same (broken) symmetries and scales. Details on the dual microscopic theories could be provided by
embedding our model in full-fledged string or M-theory constructions.
We will consider embeddings within Kaluza-Klein reductions of eleven or ten dimensional supergravities
as well as within brane-anti brane setups.  In both cases we find indications that a consistent
top-down realization of our model requires the addition of at least another non-trivial real scalar in
the gravity action.
\subsection*{Organization of the paper and main results}

This paper is organized as follows. In Section \ref{cgrev} we provide a short review of
the main features of unbalanced Fermi systems at weak coupling. In
Section \ref{modellino} we present our gravity model and write down
the corresponding equations of motion with appropriate ansatze for
the gravity fields. 

In Section
\ref{normal} we describe the $U(1)^2$-charged Reissner-Nordstrom
$AdS$ solution corresponding to the normal non-superconducting phase,
where the scalar field is zero. We discuss under which
conditions this phase could remain stable at zero temperature under
fluctuations of the scalar field envisaging the possibility of
Chandrasekhar-Clogston-like bounds depending on the choices of the
parameters (equation (\ref{semi})). 

In Section \ref{superc} we present the results of the
numerical analysis of the coupled differential equations
 when the scalar field has a non-trivial profile, i.e. in
the superconducting phase.
We thus describe the behavior of the condensate as a function of the temperature and the
$(T_c,\delta\mu)$ phase diagram (figure \ref{phasediagram}), where $T_c$ is found to be a never vanishing decreasing function of $\delta\mu/\mu$ . We also briefly comment on the (non) occurrence of LOFF phases within
our model focusing on the large charge $q\gg1$ limit. 

In Section \ref{conduct} we study the
conductivity matrix at various values of $(T,\delta\mu)$ and the comparison with weakly coupled spintronics.
Starting from our general formulas (\ref{alphaT}), (\ref{betaT}), (\ref{kappa}), we provide some precise relations among the various conductivities in the normal phase (formula (\ref{superrelazione})).
Interestingly, in the superconducting phase we find that in our strongly coupled model the DC conductivity related to the ``spin'' is enhanced, analogously to the ``electric'' one.
Moreover, the ``pseudo-gap frequency'' $\omega_{\text{gap}}$ in the superconducting phase is found to be non-linearly decreasing with $\delta\mu/\mu$; the same happens for $\omega_{\text{gap}}/T_c$.

In Section \ref{embeddings} we comment on possible string or M-theory
embeddings, providing evidence that an extra uncharged scalar is probably needed for the purpose, even in the unbalanced normal phase. It would be relevant to understand the physical meaning of this field and the role it could play within possible holographic realizations of ferromagnetic phases with magnon order parameters. We end up with appendices containing some review material and technical details.

\section{Unbalanced supercondutors}
\begin{figure}[t]
\centering
\includegraphics[scale=0.7]{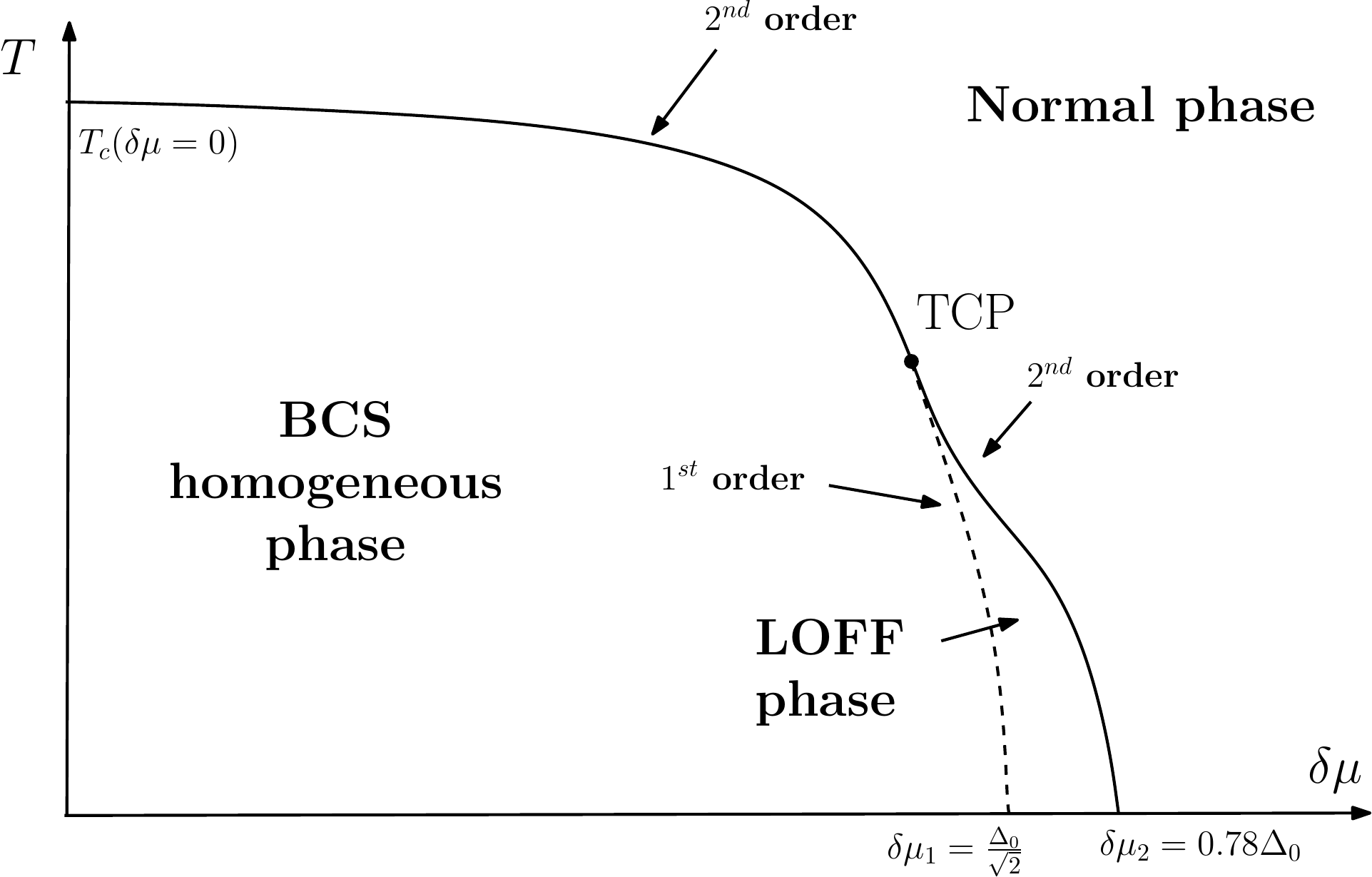}
\caption{\label{figura7} The $(T, \delta\mu)$ phase diagram for weakly coupled unbalanced superconductors.}
\end{figure}

\setcounter{equation}{0} \label{cgrev}
At weak coupling, where BCS theory can be applied, the physical properties of unbalanced
superconductors as functions of the chemical potential mismatch $\delta\mu$ are well known. The
$(T,\delta\mu)$ phase diagram is sketched in Figure \ref{figura7}.
At zero temperature, the imbalance produces a separation of the Fermi surfaces of the two fermionic
species, so that their condensation into Cooper pairs is suppressed as $\delta\mu$ is increased. As it
was shown by Chandrasekhar and Clogston \cite{cc} (whose arguments we review in appendix
\ref{chandrase}), the BCS superconducting phase becomes energetically unfavored whenever $\delta\mu$
exceeds a limiting value $\delta\mu_1=\Delta_0/\sqrt{2}$, where $\Delta_0$ is the BCS gap parameter at
$T=\delta\mu=0$. If no other phases would develop, at $\delta\mu=\delta\mu_{1}$ there would be a first order phase transition (with the gap jumping discontinuously from zero to $\Delta_0$) from the normal to the superconducting BCS phase.

Increasing the temperature above zero, a line of first order phase transitions emerges from the point
$(0, \delta\mu_{1})$ in the $(T,\delta\mu)$ phase diagram. This line joins, at a critical point, with
a line of second order transitions departing from $(T_c^0, 0)$ (at zero chemical potential mismatch,
the phase transition is the standard BCS second order one). The critical temperature $T_c$ below which
the system is superconducting is a monotonically decreasing function of $\delta\mu$. But this is not
the end of the story.

In 1964 Larkin and Ovchinnikov and independently Fulde and Ferrel \cite{loff} showed that at $T=0$,
there is a window of values $\delta\mu_{1}\le \delta\mu \le \delta\mu_{2}$ for which the energetically
favored phase is an inhomogeneous superconducting (LOFF) phase, with Cooper pairs having a non-zero
total momentum $\vec k$ (see also \cite{comb} and \cite{combmora} for reviews). The modulus $k=|\vec
k|$ is fixed by free energy minimization with respect to $k$ (this amounts to setting to zero the
superconducting current) and it results to be proportional to $\delta\mu$. Its direction is chosen
spontaneously. In general there could be crystalline phases where the fermionic condensate is a
combination of wave functions with differently directed $\vec k_i$ vectors. The simplest (FF and LO)
ansatze amount to just assume a plane wave or two-plane wave (sine or cosine) form for the condensate.
In these cases, the LOFF-to-normal phase transition at zero temperature is second order and
$\delta\mu_{2}$ is just slightly bigger than $\delta\mu_{1}$. In general, the width of the LOFF window
in the phase diagram depends on the geometry of the crystal structure in momentum space.

At finite temperature, a line of second order phase transitions departs from the point
$(0,\delta\mu_{2})$ in the phase diagram, and ends up in the previously mentioned critical point which
thus becomes a tricritical (TCP) one. 

\subsection{Unconventional superconductors}
In metallic superconductors, the Zeeman coupling with an external magnetic field (which induces a
chemical potential imbalance) is usually negligible with respect to the orbital coupling. The latter can be
naturally reduced in layered setups, such as the high $T_c$ cuprates or iron pnictides, by taking the
external magnetic field directed along the layers. These unconventional compounds are thus promising
candidates for looking at LOFF-like effects, though their non-BCS nature makes the previously
mentioned theoretical analysis not reliable. The same remark holds for unbalanced superconducting
quark matter \cite{coldqcd} whenever perturbation theory cannot be applied (e.g. for the estimated quark densities
in the core of neutron stars). In this case, one can get some insight by means of e.g.
Nambu-Jona-Lasinio (NJL)-like effective field theories, while lattice computations suffering from the
so-called sign problem are not well suited for finite density setups.

A promising setup where the LOFF effect can be experimentally investigated in various ranges of the
``coupling'', is that of trapped cold Fermi gases (see \cite{stringari} for a general review). The latter 
experience a crossover transition between a weakly coupled BCS and a strongly coupled BEC
(Bose-Einstein Condensate) phase as some parameter is varied. Experiments \cite{exp1} on cold atomic
gases with unbalanced populations seem to suggest that a Phase Separation (PS) scenario
 \cite{Bedaque} (with an homogeneous superfluid core and a normal surrounding shell) is realized
 instead of the LOFF one. However, a clear experimental evidence is still lacking,
 and the possibility of a LOFF phase emerging in some range of the coupling cannot be discarded.
 This issue has been theoretically investigated (in the mean field approximation) in e.g. \cite{Mannarelli}
 by means of NJL models with four-fermion interactions. It turns out that, at
fixed total density, when the coupling is tuned from weak to strong (i.e. in the BEC phase) the
one-plane-wave LOFF window shrinks to a point and then disappears as the coupling is increased. At
strong coupling, the unique homogeneous superfluid phase exhibits one gapless mode and has a second
order transition to the normal phase. We will see that most of these properties are reproduced by our
holographic model.


\section{Unbalanced holographic superconductors}
\setcounter{equation}{0} \label{modellino} As we have recalled in the Introduction, in the case of
metallic superconductors, a chemical potential mismatch between spin ``up'' and ``down'' species can be
induced by the Zeeman effect i.e. by the interaction term ${\cal H}_I=\bar\Psi \gamma^0
H_z\mu_B\sigma_3 \Psi$ in the Hamiltonian, between e.g. paramagnetic impurities, modeled by an
external magnetic field $H_z$ and the spin of the fermions. Here $\mu_B$ is the Bohr magneton and
$\sigma_3 = {\rm diag}(1,-1)$. The effective chemical potential mismatch is given in this case by
$\delta\mu=H_z\mu_B$. At low energy, the latter can be read as the temporal component of an effective
$U(1)_B$ spin vector field $B_\mu$ (see also \cite{iqbal}). The two fermionic species have opposite ``charges'' with respect to
$U(1)_B$, hence the up-down Cooper condensate (charged under an electromagnetic $U(1)_A$ vector field
whose temporal component effectively provides the mean chemical potential $\mu$) has zero total spin
charge. This simple picture lies at the basis of our effective holographic model and can be applied to
more general unbalanced setups (e.g. to QCD-like ones).

The simplest gravity model in $3+1$ dimensions aiming at implementing holographically the main
features of the quantum critical region of s-wave unbalanced unconventional superconductors in $2+1$
dimensions is thus described by the following action (see appendix \ref{general} for a generalization
of this setting to any $d>2$)
\begin{equation}\label{gravlag4}
S=\frac{1}{2\kappa^{2}_{4}}\int dx^4 \sqrt{-g}\left[ \mathcal{R}+ \frac{6}{L^2}-
\frac{1}{4}F_{ab}F^{ab} - \frac{1}{4}Y_{ab}Y^{ab} - V(|\psi|)-|\partial \psi-iqA\psi|^{2}\right]\,,
\end{equation}
which is a very simple extension of that introduced in \cite{hhh1,hhh2} for the balanced case. The Maxwell field $A_a$ (resp. $B_a$) with field strength $F=dA$ (resp. $Y=dB$) is the holographic dual of the $U(1)_A$ ``charge'' (resp. $U(1)_B$ ``spin'') current of the $2+1$ dimensional field theory; when $Y=0$ the system reduces to that introduced in \cite{hhh1,hhh2}. The metric is mapped into the field theory stress tensor. Finally, the complex scalar field $\psi$ is dual to the condensate. 

The action is chosen so that it admits an $AdS_4$ solution of radius
$L$ when all the matter fields are turned off. Since we are interested in finite temperature setups we will focus on asymptotically $AdS_4$ black hole solutions. Notice that the fields in the action are dimensionless and the gauge couplings have been reabsorbed in the overall gravitational constant
$\kappa_{4}^2$. Hence  the $U(1)_A$ charge $q$ of the scalar field has  dimension
of an energy.  

The functional form of the potential $V(|\psi|)$ is not apriori constrained by the symmetries in the model. For simplicity, as in \cite{hhh1,hhh2}, we will consider a potential containing only the
mass term
\begin{equation}\label{pote}
 V(|\psi|)=m^{2}\psi^{\dagger}\psi\,.
\end{equation}
By means of standard $AdS$/CFT map, the scalar field
$\psi$ results to be dual to a charged scalar operator of dimension
\begin{equation}
 \Delta(\Delta-3)=m^{2}L^{2}\,.
\end{equation}
With the aim of describing a fermionic Cooper pair-like condensate,
 $\mathcal{O}_{\Delta}=\Psi^{T}\Psi$ with canonical dimension $\Delta=2$ , we will mainly focus
 on a particular choice for the mass of the scalar field: $m^{2}=-2/L^2$.
 Of course there is no reason to believe that this will be actually the dimension of the order
 parameter driving superconductivity at strong coupling, so we will also consider other mass values at $T=0$.
 In any case the mass parameter will be taken to be above the the Breitenlohner-Freedman (BF) bound in $AdS_4$,
\be
 m^{2}L^2\geq -\frac94\,.
 \ee

\subsection{Ansatz and equations of motion}
From the action (\ref{gravlag4}) one obtains the following equations
of motion: Einstein's equations 
\begin{equation}\label{eom1}
R_{ab}-\frac{g_{ab} R}{2} -\frac{3 g_{ab}}{L^2}=-\frac{1}{2}T_{ab}\,,
\end{equation}
where
\begin{eqnarray}\label{stressenergy}
 T_{ab}&=& -
F_{ac}F_{b}^{c}-Y_{ac}Y_{b}^{c}+ \frac{1}{4}g_{ab}F_{cd}F^{cd}+ \frac{1}{4}g_{ab}Y_{cd}Y^{cd}\nonumber\\
 && + g_{ab}V(|\psi|)+ g_{ab} |\partial \psi-iqA\psi|^{2} \nonumber\\
&&-[(\partial_{a}\psi-iqA_{a}\psi)
(\partial_{b}\psi^{\dagger}+iqA_{b}\psi^{\dagger})+(a\leftrightarrow b)]\,,
\end{eqnarray}
\noindent Maxwell's equations for the $A_a$ field
\begin{equation}\label{eom2}
 \frac{1}{\sqrt{-g}}\partial_{a}(\sqrt{-g}g^{ab}g^{ce}F_{bc})=iqg^{ec}[\psi^{\dagger }(\partial_{c}\psi-iqA_{c}\psi)-
\psi(\partial_{c}\psi^{\dagger }+iqA_{c}\psi^{\dagger})]\,,
\end{equation}
 \noindent the scalar equation
\begin{equation}\label{eom3}
-\frac{1}{\sqrt{-g}}\partial_{a}[\sqrt{-g}(\partial_{b}\psi-iqA_{b}\psi)g^{ab}]
+iqg^{ab}A_{b}(\partial_{a}\psi-iqA_{a}\psi)+\frac{1}{2}\frac{\psi}{|\psi |}V^\prime (|\psi|)=0\,,
\end{equation}
\noindent and  Maxwell's equations for the $B_a$ field
\begin{equation}\label{eom5}
 \frac{1}{\sqrt{-g}}\partial_{a}(\sqrt{-g}g^{ab}g^{ce}Y_{bc})=0\,.
\end{equation}
\noindent Let us now look for static asymptotically $AdS$ black hole
solutions of the previous equations. These solutions will be dual to the the equilibrium phases of the dual quantum field theory.

For our purposes the most general ansatz for the spacetime metric
is
\begin{equation}\label{hairyblackhole}
 ds^{2}=-g(r)e^{-\chi(r)}dt^{2}+\frac{r^{2}}{L^{2}}(dx^{2}+dy^{2})+\frac{dr^{2}}{g(r)}\,,
\end{equation}
 \noindent  together with a homogeneous ansatz for the fields
\begin{equation}\label{homogeneous}
\psi=\psi(r), \quad A_a dx^a=\phi(r)dt, \quad B_a dx^a=v(r)dt\,.
\end{equation}
We will focus on  black hole solutions, with a horizon at $r=r_H$
where $g(r_H)=0$. The temperature of such black holes is given by
\begin{equation}\label{tempgen}
T=\frac{g^\prime(r_H)e^{-\chi(r_H)/2}}{4\pi}\,.
\end{equation}
\noindent Using one of Maxwell's equations we can safely choose $\psi$
 to be real. The scalar equation becomes
\begin{equation}\label{equazione1}
 \psi^{\prime\prime}+\psi^{\prime}\biggl(\frac{g^\prime}{g}+\frac{2}{r}-\frac{\chi^\prime}{2}\biggr)
-\frac{V^\prime(\psi)}{2g}+\frac{e^{\chi}q^{2}\phi^{2}\psi}{g^{2}}=0\,,
\end{equation}
\noindent Maxwell's equations for the $\phi$ field become
\begin{equation}\label{equazione2}
 \phi^{\prime\prime}+\phi^{\prime}\biggl(\frac{2}{r}+\frac{\chi^\prime}{2}\biggr)-\frac{2q^{2}\psi^{2}}{g}\phi
 =0\,,
\end{equation}
\noindent the independent component of Einstein's equations
yield
\begin{equation}\label{equazione3}
 \frac{1}{2}\psi^{\prime 2}+\frac{e^{\chi}(\phi^{\prime 2}+v^{\prime2})}{4g}+\frac{g^\prime}{gr}+
\frac{1}{r^{2}}-\frac{3}{gL^{2}}+\frac{V(\psi)}{2g}+\frac{e^{\chi}q^{2}\psi^{2}\phi^{2}}{2g^{2}}=0\,,
\end{equation}
 \begin{equation}\label{equazione4}
 \chi^\prime+r\psi^{\prime2}+r\frac{e^{\chi}q^{2}\phi^{2}\psi^{2}}{g^{2}}=0\,,
\end{equation}
\noindent Maxwell's equation for the $v$ field becomes
\begin{equation}\label{equazione5}
 v^{\prime\prime} +v^{\prime}\biggl(\frac{2}{r}+\frac{\chi^\prime}{2}\biggr)=0\,.
\end{equation}
When $v(r)=0$ these equations
reduce to those found in \cite{hhh2}. As already mentioned we will  specialize to the case where the
scalar potential contains only the mass term. In appendix \ref{general} we have
reported the same equations of motion for a general dimension $d$ of
the spacetime. In the following we will work in units $L=1$,
$2\kappa_4^2=1$.

\subsection{Boundary conditions}
\label{bdrycon} In order to find the solution to equations (\ref{equazione1}-\ref{equazione5}) one
must impose two suitable boundary conditions: one in the interior of
the spacetime at $r=r_H$ and one at the conformal boundary
$r=\infty$, where we require $AdS$ asymptotics. The analysis here is the
standard  one \cite{hhh1,hhh2}. At the horizon, both
$g(r)$ and the temporal components of the gauge fields should be vanishing \cite{hhh1}.
Hence we will require
\begin{equation}\label{IRcond}
 \phi(r_H)=v(r_H)=g(r_H)=0\,,\quad\mathrm{and} \quad  \psi(r_H),\chi(r_H) \enskip\mathrm{constants}.
\end{equation}
The series expansions  of the fields out of the horizon $r_H$,
implementing the above boundary conditions, are the following
\begin{eqnarray}\label{IRseries}
&&\phi_{H}(r) = \phi_{H1} (r-r_H) + \phi_{H2}(r-r_H)^2 +\dots ,\label{crocco1} \\
&&\psi_{H}(r) = \psi_{H0} + \psi_{H1} (r-r_H) + \psi_{H2}(r-r_H)^2+ \dots ,\label{crocco2} \\
&&\chi_{H}(r) = \chi_{H0}+\chi_{H1} (r-r_H) + \chi_{H2}(r-r_H)^2+ \dots , \\
&&g_{H}(r) =  g_{H1} (r-r_H) + g_{H2}(r-r_H)^2 + \dots ,  \\
&&v_{H}(r) = v_{H1} (r-r_H) + v_{H2}(r-r_H)^2 +\dots.
\end{eqnarray}
At the conformal boundary we must impose  a leading behavior
according to the
 corresponding dual boundary operators. Choosing $m^2L^2=-2$, the scalar field should approach
 the boundary in the following way
\begin{equation}\label{psiUV}
\psi(r)=\frac{C_1}{r}+\frac{C_2}{r^2}+\dots,\quad \mathrm{as}\quad r\rightarrow\infty.
\end{equation}
With a homogeneous anstaz  (\ref{homogeneous})  $C_1$ and $C_2$ are
constants, independent on the field theory coordinates $x_{\mu}$.
Our choice of mass does not lead to non-normalizable modes
 ($-\frac{9}{4}<m^2L^2<-\frac{5}{4}$), hence we can in principle choose whether the leading or the subleading
behavior in (\ref{psiUV}) should be the source of the dual operator
$\mathcal{O}$. Since we want the condensate to arise spontaneously, we shall require either one or the other
independent parameter in (\ref{psiUV}) to vanish. Specifically we
will choose
\begin{equation}\label{c1c2}
C_{1}=0, \enskip \langle \mathcal{O}\rangle =\sqrt{2}C_{2}\,,
\end{equation}
where the factor $\sqrt{2}$ is a convenient normalization as in
\cite{hhh1}.

Vector fields at the boundary are given by
\begin{eqnarray}
\phi(r)&=&\mu-\frac{\rho}{r} +\dots \quad \mathrm{as}\quad r\rightarrow\infty\,, \label{phiUV}\\
v(r)&=&\delta\mu-\frac{\delta\rho}{r}+\dots\quad \mathrm{as}\quad r\rightarrow\infty\,,\label{vUV}
\end{eqnarray}
where $\mu$ (resp. $\rho$) and $\delta\mu$ (resp. $\delta\rho$) are
the mean chemical potential (resp. the mean charge density) and the
chemical potential mismatch (resp. the charge density mismatch) of
the dual field theory.\footnote{In this paper we shall work in the
grand-canonical ensemble with fixed chemical potentials.} The reason behind the latter identifications has been explained before: a chemical potential mismatch can be realized by turning on a chemical potential for an effective $U(1)_s$ ``spin" field, under which a Cooper-like order parameter (whose gravity dual is the scalar field $\psi$) is uncharged. The $U(1)_B$ Maxwell field (of which $v(r)$ is the electric component) is the holographic realization of such a field. Of course our gravity model just provides an effective description of the symmetries and the order parameters of the dual field theory. As such, the $U(1)_A$ and $U(1)_B$ fields that we treat as the holographic duals of the ``charge" and the ``spin" currents respectively, could be actually mapped into any couple of abelian global symmetries  of which one can be broken by a vev of a charged scalar, while the other stays unbroken. 

Notice that we work in units $L=2\kappa_4^2=1$, where the bulk fields
$A_a$, $B_a$ and the parameters $\mu$, $\delta\mu$ have mass
dimension 1, while $\psi$ is dimensionless;
 $\rho$ and $\delta\rho$ are charges per unit  volume in the $(2\!+\!1)$-dimensional field theory,
hence  have dimension $l^{-2}$; the radial coordinate $r$ has
dimension 1 in mass.

The functions in the metric ansatz should have $AdS_4$ asymptotics as in \cite{hhh2}
 \begin{eqnarray}
 g(r)&=&r^2 -\frac{\epsilon}{2r} + \dots\quad \mathrm{as}\quad r\rightarrow\infty\label{gUV}\\
 \chi(r)&=&0 +\dots\quad \mathrm{as}\quad r\rightarrow\infty\,,
 \label{chiUV}
\end{eqnarray}
where $\epsilon$ is holographically mapped to the energy density of the dual field theory.
\section{The Normal Phase}
\setcounter{equation}{0} \label{normal}
 A simple solution to the equations of
motion (\ref{equazione1}-\ref{equazione5}) corresponds to the normal
phase in the dual field theory. This is characterized by a vanishing
 vacuum expectation value of the condensate $\mathcal{O}$, corresponding to
a vanishing scalar field  $\psi=0$ in the bulk. The corresponding
background is that of a $U(1)^2$-charged Reissner-Nordstrom-$AdS_4$
black hole, with metric
\begin{eqnarray}\label{rnAdS}
ds^{2}&=&-f(r)dt^{2}+r^{2}(dx^{2}+dy^{2})+\frac{dr^{2}}{f(r)},\\
f(r)&=&r^{2}\bigg(1-\frac{r_{H}^{3}}{r^{3}}\bigg)+
\frac{\mu^{2}r_H^{2}}{4r^{2}}\bigg(1-\frac{r}{r_{H}}\bigg)
+\frac{\delta\mu^{2}r_H^{2}}{4r^{2}}\bigg(1-\frac{r}{r_{H}}\bigg)\,.\label{fr}
\end{eqnarray}
Here $r_H$ is the coordinate of the black hole outer horizon. The
gauge fields are given by
\begin{eqnarray}
\phi(r)&=&\mu\bigg(1-\frac{r_{H}}{r}\bigg)=\mu-\frac{\rho }{r}\,,\label{phiN}\\
v(r)&=&\delta\mu\bigg(1-\frac{r_{H}}{r}\bigg)=\delta\mu-\frac{\delta\rho }{r}\label{vN}\,.
\end{eqnarray}
The temperature reads
\begin{equation}\label{tempN}
T=\frac{r_H}{16\pi}\bigg(12-\frac{\mu^2+\delta\mu^2}{r_H^2}\bigg)\,,
\end{equation}
from which we get
\begin{equation}\label{rhN}
 r_H=\frac{2}{3}\pi T+\frac{1}{6}\sqrt{16\pi^2T^2+3(\mu^2+\delta\mu^2)}\,.
\end{equation}
The Gibbs  free energy density reads
\begin{equation}\label{normalfree}
\omega_n=-r_H^3\bigg( 1+\frac{(\mu^2 + \delta\mu^2)}{4r_H^2}\bigg).
\end{equation}
Notice that, due to formula (\ref{rhN}), this is a function of $T$, $\mu$
and $\delta\mu$.

The doubly charged $AdS$ black
hole solution is certainly not new in the holographic condensed matter
literature. For example, it also describes the normal phase of
unbalanced p-wave superconductors \cite{erdmeng}. As it was noticed in \cite{iqbal}, the normal phase of a model like ours can be also seen as a rough holographic realization of a ``forced''
ferromagnet, where the ``spin'' density $\delta\rho$ is supported by a
non zero value of $\delta\mu$ (and so, equivalently, by an external
magnetic field). Indeed in our case $\delta\rho=0$ if $\delta\mu=0$.
\footnote{In a real ferromagnet, instead, the spin density $\delta\rho$
is spontaneously generated. Moreover there is a non vanishing
ferromagnetic order parameter (a magnon). The holographic description
of such a setup is an interesting open issue.}

At $T=0$ the doubly charged RN-$AdS_4$ black hole becomes extremal
and (see equation (\ref{tempN})) the  horizon radius reads
\begin{equation}
 r_H^2=\frac{1}{12}(\delta\mu^2+\mu^2)\quad\mathrm{at}\quad T=0\,.
\label{rhN0}
\end{equation}
In the near-horizon limit the metric reduces to that of an
$AdS_2\times R^2$ background with $AdS_2$ radius given by
$L_{(2)}^2=L^2/6$ (see also Appendix \ref{general}).

The charge density imbalance at $T=0$ reads \be \delta\rho = \sqrt{\frac{\mu^2+\delta\mu^2}{12}}
\delta\mu\,. \ee Notice that this is zero at $\delta\mu=0$ as it happens for the normal phase at weak coupling (see appendix
\ref{chandrase}). The susceptibility imbalance (``magnetic'' susceptibility) reads thus \be \delta\chi
= \frac{\partial \delta\rho}{\partial\delta\mu}|_{\delta\mu=0} = \frac{\mu}{\sqrt{12}}\,. \ee Therefore, in
the limit $\delta\mu \ll \mu$, the Gibbs free energy density of the normal phase at $T=0$ goes at
leading order as \be \omega(\delta\mu)\approx \omega(0)-\frac12\frac{\mu}{\sqrt{12}}\delta\mu^2\,. \ee
Following the same reasonings as in \cite{cc} (see also appendix \ref{chandrase}), we can argue that,
provided a superconducting phase exists at $T=0$, \emph{and it has $\delta\rho=0$ for every $\delta\mu$},
a Chandrasekhar-Clogston bound at $T=0$ should naturally arise also within the holographic setup.

Let us now ask whether there are conditions under which, lowering
the temperature, a superconducting phase ($\psi\neq0$) might arise
with a formation of a charged condensate below a certain critical
temperature $T_c$.
\subsection{A criterion for instability}
\label{criterion}
In our model we can find a simple condition (see also \cite{iqbal,horrob}) on the external
parameters in order for the normal phase to become unstable at
$T=0$. Let us consider a fluctuation of the complex scalar field
$\psi$, charged under $U(1)_A$, around the extremal $U(1)^2$-charged
RN-$AdS$ background. Its equation of motion has the form given in
(\ref{equazione1}) with background metric given in (\ref{rnAdS})
and $\phi(r)$ given in (\ref{phiN}). The horizon radius is fixed as
in (\ref{rhN0}).

In the near-horizon limit it is easy (see \cite{horrob}) to see
that the equation for $\psi$ reduces to that of a scalar field of
mass
\begin{equation}
m^2_{\mathrm{eff}(2)}= m^2-\frac{2q^2}{\left(1+\frac{\delta\mu^2}{\mu^2}\right)}\,,
\end{equation}
on an $AdS_2$ background of squared radius $L_{(2)}^2=1/6$. The
instability of the normal phase in this limit, is thus mapped into
the requirement that the above effective mass is below the $AdS_2$
BF bound
\begin{equation}
L_{(2)}^2 m^2_{\mathrm{eff}(2)}= \frac16 m^2_{\mathrm{eff}(2)} < - \frac{1}{4}\,,
\end{equation}
which leads to
\begin{equation}\label{semi}
\left(1+\frac{\delta\mu^2}{\mu^2}\right) \left(m^2+\frac{3}{2}\right)<2q^2\,.
\end{equation}
For an analogous formula in the general $d+1$-dimensional case see appendix \ref{general} (see also \cite{iqbal}).

When $(m^2+\frac{3}{2})<0$, i.e. $m^2<-\frac{3}{2}$, the instability
occurs for every value of $\frac{\delta\mu^2}{\mu^2}$. This will
indeed be the case for $m^2=-2$. This suggests, quite surprisingly, that in
these cases a superconducting phase with non-trivial scalar profile
could be always preferred at $T=0$, no matter how large is the
chemical potential mismatch.\footnote{A similar result was found in
\cite{iqbal} studying the instability of an extremal dyonic black
hole, electrically and magnetically charged under a $U(1)$ field.} 

In the cases in which $(m^2+\frac{3}{2})>0$, instead, the normal phase will show instability when
\begin{equation}\label{cclike}
\frac{\delta\mu^2}{\mu^2}<2q^2\frac{1}{\left(m^2+\frac{3}{2}\right)}-1\,,
\end{equation}
which gives an actual bound on $\delta\mu/\mu$ provided that $4q^2>
2m^2 + 3$. The above condition resembles the Chandrasekhar-Clogston
bound of weakly interacting superconductors.

According to the comments in \cite{conflost}, we expect that a violation of the $AdS_2$ BF bound leads to a
continuous phase transition of the Berezinskii-Kosterlits-Thouless (BKT) type at $T=0$ (see also
\cite{jensen}). In BKT transitions the order parameter goes to zero exponentially instead as with the
power law behavior of second order phase transitions. Around these phase transitions there should be a
turning point in the phase diagram, with the critical temperature slowly going to zero as an external
parameter (for us $\delta\mu/\mu$) is increased. Actually, a BKT transition should become of second order 
at $T>0$. Moreover, in \cite{jensen} it has been observed that, in a holographic model, a BKT transition at $T=0$ can only occur when the theory has two control
parameters with the same dimension. This is precisely what happens in our case. Finally, notice that if
the normal-to-superconducting phase transition at $T=0$ is a continuous one (e.g. a BKT one) the
critical value of the parameter $\delta\mu/\mu$ as deduced from the BF bound in $AdS_2$ should
correspond to the critical value at which the phase transition occurs (see also analogous comments in \cite{erdmeng}).
\section{The Superconducting Phase}
\setcounter{equation}{0} \label{superc}
 If the normal phase becomes unstable at low $T$,
 we must search for another static solution to the
equations of motion (\ref{equazione1}-\ref{equazione5}) where  the
scalar field is non-zero. In the dual field theory this corresponds
to turning on a vacuum expectation value of the condensate  leading
to a spontaneous symmetry breaking of an electromagnetic symmetry
and the consequent emergence of a superconducting phase.

In the following we will discuss the results of a standard numerical analysis of the full set of
equations of motion subjected to the boundary conditions discussed in Section \ref{bdrycon}. The
analysis is mainly based on the shooting technique and it is strictly valid at $T>0$. We defer the
study of the $T=0$ case, along the lines considered in \cite{horrob}, to future work.
\subsection{The condensate}
Let us concentrate on what we can learn from the numeric
solutions. First of all let us find an expression for the
temperature as a function of the horizon values of the various
fields. From the general expression (\ref{tempgen}) and the Einstein
equation at the horizon, it follows that, for generic $m^2$
\begin{equation}\label{tronco}
T=\frac{r_H}{16\pi}\bigg[ (12 -2m^2\psi_{H0}^2)
e^{-\frac{\chi_{H0}}{2}}-\frac{1}{r_H^2}e^{\frac{\chi_{H0}}{2}}(\phi_{H1}^2+v_{H1}^2)  \bigg]\,.
\end{equation}
The critical temperature is found by setting $\langle \mathcal{O} \rangle \sim C_2=0$.\footnote{A
dimensionless combination involving $T$ and the chemical potentials is $T/(\mu^2+\delta\mu^2)^{1/2}$.}

Our numerical analysis gives rise to the following results.
For small values of the chemical potential mismatch $\delta\mu=0.01$
and for different values of the external parameter $q$ we obtain
results similar to \cite{hhh2}. A condensate arises below a certain
critical temperature $T_c$ signaling a phase transition from a
normal to a superconducting phase. The general form of these curves
is similar to the ones in BCS theory, typical of mean field theories
and second order phase transitions. The value of the condensate
depends on the charge of the bulk field $q$. However, as in
\cite{hhh2}, it is difficult to get the numerics reliably down to
very low temperatures.

\begin{figure}[t]
\centering
\includegraphics[width=120mm]{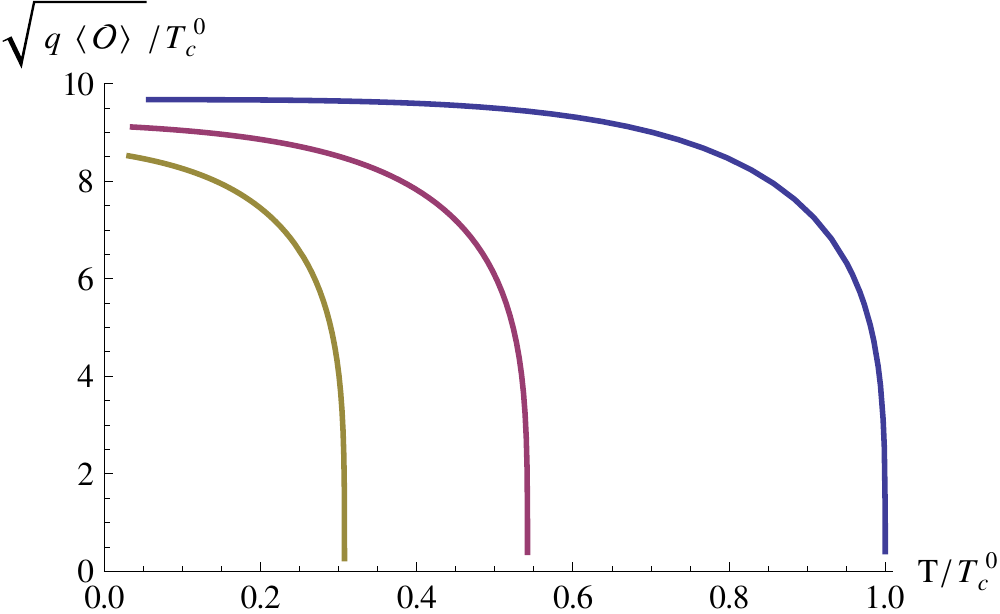}
\caption{\label{Condensato12}The value of the condensate as a function of the temperature at $\mu = 1, q = 2$.
From right to left we have $\delta\mu = 0, 1, 1.5$ and the critical temperature is
decreasing.
}
\end{figure}

\begin{figure}[t]
\centering
\includegraphics[width=120mm]{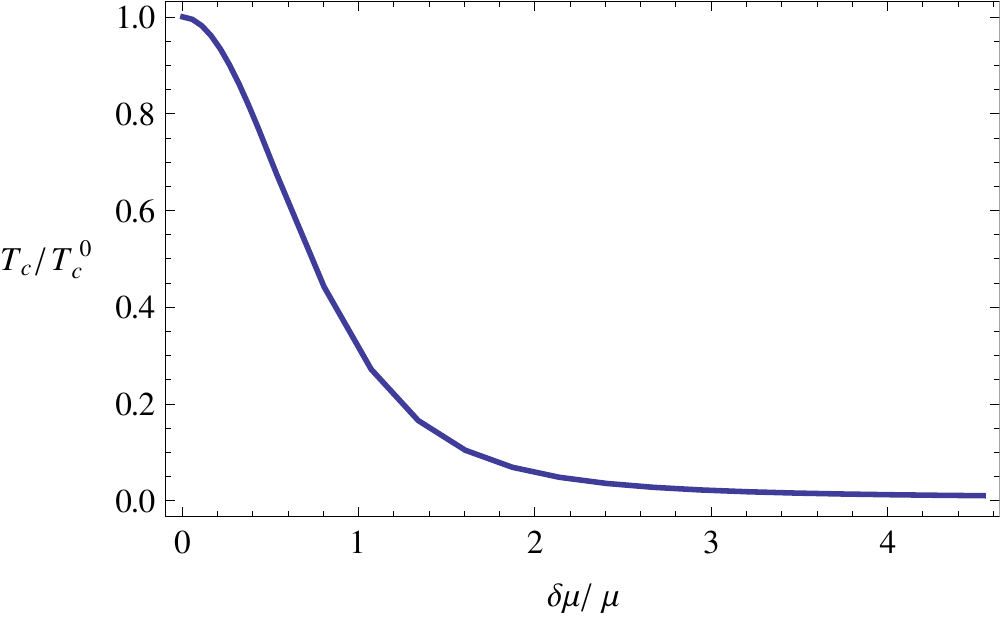}
\caption{\label{phasediagram}Second order phase transition line in the $(T_c, \delta\mu)$ plane
with $q=1$, $\mu=1.87$.
There are always values of $T_c$ below which a superconducting phase arises. }
\end{figure}

Allowing for higher values of the chemical potential mismatch
$\delta\mu$, we obtain analogous behaviors for the condensate.
Increasing the value of $\delta\mu$ (see Figure \ref{Condensato12})
we obtain a decreasing value of the critical temperature. The phase
transition is always second order.
The most interesting result is the plot of the critical temperature
normalized to $T_c^0$ (the critical temperature at zero chemical
potential mismatch), against $\delta\mu/\mu$. The second
order phase transition  at zero chemical potential mismatch develops
inside the $(T_c, \delta\mu)$ phase diagram. As it is shown in Figure
\ref{phasediagram}, the critical temperature decreases with
$\delta\mu/\mu$, a qualitative feature which we have seen
also in the weakly coupled case.
However, differently from the weakly coupled case, there is no finite value of $\delta\mu/\mu$ for
which $T_c=0$.\footnote{A similar phase diagram appears in \cite{Nakano:2008xc} in holographic
superconductors in the presence of an external magnetic field.} Hence, there is no sign of a
Chandrasekhar-Clogston bound at zero temperature. This result matches with the expectations coming
from the formula (\ref{semi}), which actually suggested the absence of such a bound for $m^2=-2$.
However, it should be desirable to refine our numerics around $T=0$ as done in \cite{horrob} to
definitely confirm this conclusion. In any case, we believe that it is unlikely that the curve in
Figure \ref{phasediagram} will suddenly drop to zero with another flex. The phase transition we find
is always second order as we have also checked by a standard holographic computation of the free
energy along the same lines as in \cite{hhh2}.\footnote{Notice that, as we have previously argued, the
phase diagram could change for different choices of the mass parameters in our model. We defer this
analysis to future works.} Together with the absence of a Chandrasekhar-Clogston bound, this leads us
to argue that a LOFF phase is unlikely to develop.
\subsection{Comments on the LOFF phase}
Let us see whether the occurrence of inhomogeneous phases is actually forbidden in our model. For
simplicity, we will limit our analysis to the so-called ``probe'' approximation as first considered in
\cite{hhh1}. It consists in rescaling the scalar as well as the $A_a$ gauge field by the charge $1/q$
and taking the $q\gg1$ limit. In this way the backreaction of these fields on the metric can be
neglected. This limit could only capture the physics at temperatures close to the critical temperature
$T_c$, where the condensate (hence the field $\psi$) is actually small. In particular, it could be reliable around a tricritical point, in case this is displayed by the $(T,\delta\mu)$ phase diagram.

Looking for LOFF-like phases in this limit corresponds to looking for gravity solutions (on a fixed
background) where the scalar field spontaneously acquires a dependence on e.g. one of the spatial
directions. In particular we will focus on complex one-plane wave solutions and on real sinusoidal
ones.\footnote{Inhomogeneous phases of this kind in holographic p-wave superconductors have been
considered in, e.g. \cite{ooguri,stripped}.} In our case the fixed background is given by a
$U(1)_B$-charged RN-$AdS$ black hole
\begin{eqnarray}
ds^{2}&=&-f(r)dt^{2}+r^{2}(dx^{2}+dy^{2})+\frac{dr^{2}}{f(r)}\,,\label{procione1}\\
f(r)&=&r^{2}(1-\frac{r_{H}^{3}}{r^{3}})+\frac{\delta\mu^{2}r_H^{2}}{4r^{2}}(1-\frac{r}{r_{H}})\,,\label{procione2}\\
v_{t}&=&\delta\mu(1-\frac{r_{H}}{r})=\delta\mu-\frac{\delta\rho }{r}\,.\label{procione3}
\end{eqnarray}
Let us now consider a simple single plane wave inhomogeneous ansatz
\be
\psi(r,x)=\Psi(r)e^{-ixk}\,,\quad A=A_t(r)dt + A_x(r)dt\,.
\ee
Inserting the above ansatz in the equations of motion (\ref{eom2}) and
(\ref{eom3}) on the fixed $U(1)_B$-charged RN-$AdS$ background
(\ref{procione1}), (\ref{procione2}), (\ref{procione3}), one gets
Maxwell's equations
\begin{eqnarray}
\partial_r^2A_t+\frac{2}{r}\partial_rA_t-\frac{2\Psi^2}{f}A_t=0\,,\\
 \partial_r^2A_x+\frac{f^\prime}{f}\partial_r A_x-\frac{2\Psi^2}{f}(k+A_x)=0\,,
\end{eqnarray}
and the scalar equation
\begin{equation}
\Psi^{\prime\prime}+\Psi^\prime\bigg( \frac{2}{r}+\frac{f^\prime}{f}\bigg)+
\Psi\bigg( \frac{A_t^2}{f^2}+\frac{2}{f}-\frac{(k+A_x)^2}{r^2f}\bigg)=0\,.
\end{equation}
These equations always admit a trivial solution with $A_x=-k$ which is just related to the $k=0$ one
by bulk gauge transformations. Both solutions thus correspond to the same zero-current homogeneous
phase. In order to find non-trivial solutions corresponding to an absolute equilibrium state with zero superfluid current\footnote{This is a
necessary condition for having a LOFF ground state \cite{loff} and it marks a difference w.r.t. the
standard superfluid phases with non zero superfluid velocity, see e.g. \cite{superflu,bert}.} we could
require the Maxwell field $A_x(r)$ to have UV boundary conditions such that \be
A_x(r\rightarrow\infty)\approx -k + \frac{J}{r}\,, \quad{\rm with}\quad J=0\,. \label{bdrax} \ee The
large-$r$ asymptotics for $\Psi$ and $A_t$ are taken as in the homogeneous case. From the above
equations it is easy to realize that non-trivial solutions for $A_x$ satisfying the boundary condition
(\ref{bdrax}) are not admitted. This is true for generic values of $m^2$, for which the scalar field
$\Psi(r)$, dual to the same kind of operator we have considered above, has to go like $\Psi\approx C
r^{-\lambda}$ at large $r$ (here $\lambda= (3+\sqrt{9+4m^2})/2$ is the operator UV dimension). The
same conclusion is reached starting from a more general ansatz in which $A_t=A_t(x,r)$ and
$A_x=A_x(x,r)$, with the same large $r$ asymptotics as in (\ref{bdrax}). Maxwell's equations in this
case imply separability $\partial_x\partial_r A_x =0$ and consistency forces $A_x$ and $A_t$ to loose
their dependence on $x$, thus reducing the setup to the one above.

It is also possible to show that two-wave real condensates going like $\psi=\Psi(r)\cos(kx)$ are excluded already at the level of the equations of motion. 

All in all, this analysis seems to exclude the possibility that a LOFF-like phase, and thus a related tricritical point in the $(T,\delta\mu)$ phase diagram, can be displayed by our model.

\section{Conductivities: holographic spintronics}
\setcounter{equation}{0} \label{conduct}

In this section we present the study of the conductivities of the system as functions of the frequency of the applied external field perturbations.
In this paper we limit the analysis to zero-momentum perturbations.
Thanks to the rotational invariance of the theory in the $x-y$ plane, it is sufficient to consider the
conductivities in one direction, say along $x$.
According to the $AdS$/CFT dictionary, the calculation is performed through the study of the response of the gravity system to
vector perturbations in the $x$ direction.
The holographic computation, which is fairly standard, is described in Section \ref{conto_conductivity}; the un-interested reader can safely skip it (apart from formulas (\ref{alphaT}), (\ref{betaT}), (\ref{kappa})) and go to the results in Sections \ref{res_normal} and \ref{res_sucond}.

\subsection{The conductivity matrix}

The system at hands includes two $U(1)$ vector fields describing two currents of the dual boundary
theory. As mentioned in the Introduction, it represents a minimal holographic description of the ``two-current model'' \cite{mott,fert1,vanson} which lies at the roots of the theoretical study of
spintronics.
The two-current model is based on the observation by Mott \cite{mott} that at low temperature the two
currents of spin ``up'' and ``down'' electrons can be considered as two separate entities, with two
corresponding conductivities which are in principle different. The system is typically a ferromagnetic
metal with a non zero spin polarization $\delta\rho$.

In such case, an applied external electric field $E^c$ (providing an ``electric motive force'') does
not generate only an electric current $J^c$ with conductivity $\sigma_{cc}$, but can also generate a
net spin current $J^s$. The corresponding conductivity is sometimes called (somehow asymmetrically)
the ``spin conductivity'' $\sigma_{cs}$.
Moreover, this effect is reciprocal: a ``spin motive force'' generated by an external applied field
$E^s$ (essentially, a dynamical gradient of population imbalance,\footnote{This can be generated by
``dynamical magnetic textures'', e.g. dynamical magnetizations in ferromagnetic conductors. Typical
potential differences are generated by appropriately engineered sequences of layers of materials with
different magnetic properties. Thus, the ``spin motive force'' is generated perpendicularly to the
layers. In our gravity dual model, we instead consider the ``spin motive force'' in the plane; for our
purposes the difference of the two settings is irrelevant.} $\nabla \delta\mu$), on top of creating a
spin current $J^s$ with conductivity $\sigma_{ss}$ (we call this the ``spin-spin conductivity''),
induces an electric current $J^c$ as well, with conductivity $\sigma_{sc}$ \cite{vanson}; the latter
is precisely equal to the spin conductivity $\sigma_{cs}$ in time-reversal invariant settings. We will
not consider dissipative effects bringing to spin-relaxation in this paper.

The electric and spin motive forces can be described by means of two $U(1)$ gauge fields: the
electro-magnetic one and the spin one; the latter is an effective gauge boson.
In this paper we have presented a gravity theory which precisely encodes the minimal ingredients of a
macroscopic description of the two-current model: a charged environment, i.e. the charged black hole,
with two vector operators, dual to the two gauge fields $A_a$ and $B_a$. The two gauge fields are not
directly coupled in the gravity Lagrangian, but their fluctuations are coupled via their coupling with
the metric. As a result, the two dual currents are coupled through their mixing with the
energy-momentum tensor operator of the field theory.
Obviously enough, there is a mixing with the heat current as well.

Thus, we would like to stress the fact that from the dual gravitational point of view, the mixing of the two current operators, which causes the crucial existence of the ``spin conductivity'' $\sigma_{cs}=\sigma_{sc}$, is completely straightforward and universal.
In fact, both currents bring some momentum, thus they source the momentum operator $T_{tx}$; the latter is dual to the metric component $g_{tx}$ which, being a vector perturbation of the metric, mixes with both gravity vector fields, coupling the dual operators.
In other words, this phenomenon is not very sensitive to the details of the gravity Lagrangian used to describe the two-current model: generally, a gravity theory with two conserved $U(1)$'s will provide a non-zero dual ``spin conductivity''.

Furthermore, the gravity theory at hands includes a charged operator (under the ``electric charge'') which condenses at sufficiently small temperature.
Thus, the system describes both the normal phase of a two-current model, and a superconducting phase thereof.
It must also be remembered that the gravity theory provides a dual description of strongly interacting microscopic degrees of freedom,
as opposed to the usual weakly interacting fermion systems of the standard spintronics literature.

Actually, the gravity description does not rely on the specific microscopic origin of the two $U(1)$ currents in terms of charge and spin, although it describes the same basic features.
In this sense, it is more ``universal'' and could be possibly applied to other microscopic theories, such as QCD-like ones.
Thus, we chose to call $J^A, E^A, \sigma_A$ the current, external field and conductivity associated to the $U(1)_A$ under which the scalar operator is charged; these are the quantities which we called ``electric'' above (i.e. electric current $J^c$, field $E^c$, conductivity $\sigma_{cc}$).
Analogously, we call $J^B, E^B, \sigma_B$ the quantities associated to the $U(1)_B$ under which the scalar operator is un-charged; above we referred to them in relation to the ``spin'' (i.e. they are $J^s$, $E^s$, $\sigma_{ss}$ in the case of electron spin unbalance).
Finally, we name $\gamma$ the mixed conductivity ($\sigma_{cs}=\sigma_{sc}$ above).
Nevertheless, in the discussion below we will often indulge with the ``electric/spin'' terminology for a (hopefully) clearer presentation.

According to the discussion above, all the conductivities are included in the general non-diagonal matrix (see e.g. \cite{vanson} and \cite{hhh2} for the notation)
\begin{eqnarray}\label{matrix}
\begin{pmatrix}  J^A \\ Q \\ J^B \end{pmatrix} = \begin{pmatrix}  \sigma_A & \alpha T & \gamma \\ \alpha T &
\kappa T & \beta T \\ \gamma & \beta T & \sigma_B \end{pmatrix} \cdot \begin{pmatrix}  E^A \\ -\frac{\nabla T}{T} \\ E^B \end{pmatrix} \, .
\end{eqnarray}
In this matrix, $\sigma_A, \sigma_B$ are the diagonal conductivities associated to the two $U(1)$'s.
In the language used above, the former is the standard electric conductivity, measuring the proportionality between an applied external electric field $E^A$ and a generated electric current $J^A$.
Analogously, the ``spin-spin'' conductivity $\sigma_B$ is the proportionality coefficient between a gradient in the imbalance chemical potential $\delta\mu$ (which we called $E^B$ above to underline the similarity with $E^A$), and the generated spin current $J^B$.
The third diagonal entry in (\ref{matrix}), $ \kappa $, is the thermal conductivity, i.e. the proportionality between the thermal gradient $-\frac{\nabla T}{T}$ and the heat current $Q=T_{tx}-\mu J^A - \delta\mu J^B$.

Moreover, $\alpha$ and $\beta$ are the thermo-electric and ``thermo-spin''
conductivities;\footnote{$\beta$ is sometimes called the ``thermo-magnetic'' conductivity
\cite{vanson}.} they are associated to the transport of heat in the presence of an electric, or spin,
potential even without a temperature gradient $\nabla T$. Finally, $\gamma$ measures e.g. the spin
current $J^B$ generated by an applied electric potential $E^A$ even without an applied external field
$E^B$.

The matrix (\ref{matrix}) is symmetric because of time-reversal symmetry \cite{vanson,hhh2}.

\subsection{Holographic calculation of the conductivities}\label{conto_conductivity}
We follow closely the presentation in \cite{hhh1,hhh2} throughout all the present section, underlining the novelties of the unbalanced case. In the holographic approach, the fluctuations of the fields $A_x, B_x, g_{tx}$ at the $AdS$ boundary, act as sources for the currents $J_x^A, J_x^B$ and the stress energy tensor component $T_{tx}$. Conductivity is a transport phenomenon, hence it requires a real
time description. Since the black hole solutions we consider are
classical, we must require in-going boundary conditions for the
fields $A_x, B_x, g_{tx}$ at the horizon. Let us take a simple $e^{-i\omega t}$
time dependence for the fluctuations and consider the related linearized Einstein and Maxwell equations on the background
\begin{equation}\label{maxa}
 A_x'' + \left(\frac{g'}{g}-\frac{\chi'}{2}\right) A'_x + \left(\frac{\omega^2}{g^2}e^\chi - \frac{2q^2\psi^2}{g}\right) A_x =
 \frac{\phi'}{g} e^\chi \left(-g'_{tx} + \frac{2}{r} g_{tx}\right)\, ,
\end{equation}
\begin{equation}\label{maxb}
 B_x'' + \left(\frac{g'}{g}-\frac{\chi'}{2}\right) B'_x + \frac{\omega^2}{g^2}e^\chi B_x =
 \frac{v'}{g} e^\chi \left(-g'_{tx} + \frac{2}{r} g_{tx}\right)\, ,
\end{equation}
\begin{equation}\label{grav}
 g'_{tx} - \frac{2}{r}g_{tx} + \phi' A_x + v' B_x= 0\, .
\end{equation}
The prime represents the derivative with respect to the bulk radial coordinate $r$.
Substituting (\ref{grav}) into (\ref{maxa}) and (\ref{maxb}) we obtain
\begin{equation}\label{eqA}
 A_x'' + \left(\frac{g'}{g} - \frac{\chi'}{2}\right) A_x'
 + \left( \frac{\omega^2}{g^2}e^{\chi} - \frac{2 q^2 \psi^2}{g} \right) A_x
 - \frac{\phi'}{g} e^{\chi} \left(B_x v' + A_x \phi' \right) = 0\, ,
\end{equation}
\begin{equation}\label{eqB}
 B_x'' + \left(\frac{g'}{g} - \frac{\chi'}{2}\right) B_x'
 + \frac{\omega^2}{g^2}e^{\chi} B_x
 - \frac{v'}{g} e^{\chi} \left(B_x v' + A_x \phi' \right) = 0\, .
\end{equation}
In this way we can deal with two equations in which the metric fluctuations do not appear. Notice that
the substitution has lead to a system of linear differential equations in which the two gauge fields $A_x$ and $B_x$ are
mixed. It is important to underline the r\^ole of the metric in such mixing: indeed, in the probe
approximation, no coupling between the different gauge fields occurs.

In order to solve (\ref{eqA}) and (\ref{eqB}), we assume the following near-horizon behavior ansatz
for the fluctuation functions\footnote{It can be checked that in order to have a non-trivial solution of the equations around the horizon, the frequencies of the two modes must be equal. Moreover, inspection of the behavior of the equations near the horizon dictates that the exponential coefficients ``$a$'' of the leading terms of the two modes must be equal too.}
\begin{eqnarray}
 & A_x(r) & = \left(1-\frac{r_H}{r}\right)^{i a \omega} \left[ a_0 + a_1 \left(1-\frac{r_H}{r}\right) + ... \right]\, ,\\ \label{asintoticiB}
 & B_x(r) & = \left(1-\frac{r_H}{r}\right)^{i a \omega} \left[ b_0 + b_1 \left(1-\frac{r_H}{r}\right) + ... \right]\,.
\end{eqnarray}
The IR solution depends on two integration constants, $a_0, b_0$.
The overall scaling symmetry $(A_x \rightarrow \lambda A_x, B_x \rightarrow \lambda B_x)$ of equations (\ref{eqA}), (\ref{eqB}) allows one to fix e.g. $a_0=1$.

In the UV, the behavior of the fields is
\begin{eqnarray}\label{boufie}
 A_x(r) &= A_x^{(0)} + \frac{1}{r} A_x^{(1)} + ... \, ,\\
 B_x(r) &= B_x^{(0)} + \frac{1}{r} B_x^{(1)} + ... \, ,\\
 g_{tx}(r) &= r^2 g_{tx}^{(0)} - \frac{1}{r} g_{tx}^{(1)} + ...\, .
\end{eqnarray}
With this notation, the solution for $g_{tx}$ in equation (\ref{grav}) can be expressed as
\begin{equation}
g_{tx} = r^2 \left( g_{tx}^{(0)} + \int_r^\infty \frac{\phi' A_x + v' B_x}{r^2} \right)\, ,
\end{equation}
so that
\begin{equation}\label{guno}
g_{tx}^{(1)} = \frac{\rho}{3}A_x^{(0)} + \frac{\delta\rho}{3}B_x^{(0)}\, .
\end{equation}
Moreover, from linearity and symmetries of (\ref{eqA}) and (\ref{eqB}) it follows that, on-shell, 
\be
A_x^{(1)} =i\omega\sigma_A A_x^{(0)} + i\omega\gamma B_x^{(0)},\qquad B_x^{(1)} =i\omega\gamma
A_x^{(0)} + i\omega\sigma_B B_x^{(0)}\,, 
\label{onshAB}
\ee 
where the reason behind the labeling of the $\omega$-dependent coefficients will be clear in a moment.

The linear response of a current $J^a$ to perturbations $\sum_b \phi_b J^b$ driven by external
sources $\phi_b$ is given in terms of retarded correlators. Formally, in momentum space,
$\langle J^a \rangle =G_{R}[J^a, J^b]\phi_b$. In our case, the retarded correlators $G_R$, which are proportional to the
conductivities, can be holographically computed using the on-shell gravity action for the dual fields
as generating functional. The on-shell action for the quadratic fluctuations of $A_x, B_x, g_{tx}$ can
be expressed as just a boundary term at infinity and, after performing the holographic renormalization
procedure, it can be reduced to \cite{hhh2}
\begin{equation}\label{resuquad}
 S_{quad} = \int d^3x \left( \frac{1}{2}A^{(0)}_x A_x^{(1)}
+ \frac{1}{2}B^{(0)}_x B_x^{(1)}
- 3 g_{tx}^{(0)} g_{tx}^{(1)} -\frac{\epsilon}{2}g_{tx}^{(0)}g_{tx}^{(0)} \right) \ ,
\end{equation}
with $A_x^{(1)}, B_x^{(1)}, g_{tx}^{(1)}$ given in (\ref{guno}) and (\ref{onshAB}).

The conductivity matrix can be thus deduced using the holographic relations
\begin{eqnarray}
&& J^A =\frac{\delta S_{quad}}{\delta A_x^{(0)}}\,, \\
&& J^B =\frac{\delta S_{quad}}{\delta B_x^{(0)}}\,, \\
&& Q =\frac{\delta S_{quad}}{\delta g_{tx}^{(0)}} - \mu J^A  - \delta\mu J^B\,,
\end{eqnarray}
provided the following formulas (see Hartnoll's and Herzog's reviews in \cite{reviewsADSCM} for
details) are employed 
\be 
E_x^A = i\omega(A_x^{(0)} + \mu g_{tx}^{(0)})\,,\quad E_x^B = i\omega(B_x^{(0)} +
\delta\mu g_{tx}^{(0)})\,,\quad -\frac{\nabla_x T}{T}=i\omega g_{tx}^{(0)}\,.
\ee 
We can thus get
\begin{eqnarray}
\sigma_A = \frac{J^{A}}{E^{A}}|_{g_{tx}^{(0)}=B_x^{(0)}=0} = - \frac{i}{\omega} \frac{A_x^{(1)}}{A_x^{(0)}}|_{g_{tx}^{(0)}=B_x^{(0)}=0}\ , \nonumber \\
\gamma =  \frac{J^{B}}{E^{A}}|_{g_{tx}^{(0)}=B_x^{(0)}=0} = - \frac{i}{\omega} \frac{B_x^{(1)}}{A_x^{(0)}}|_{g_{tx}^{(0)}=B_x^{(0)}=0}\ , \nonumber \\ 
\alpha T =  \frac{Q}{E^{A}}|_{g_{tx}^{(0)}=B^{(0)}=0} = \frac{i \rho}{\omega} - \mu \sigma_A -\delta\mu \gamma \,,
\label{alphaT}
\end{eqnarray}
as well as
\begin{eqnarray}
\sigma_B = \frac{J^{B}}{E^{B}}|_{g_{tx}^{(0)}=A_x^{(0)}=0} = - \frac{i}{\omega} \frac{B_x^{(1)}}{B_x^{(0)}}|_{g_{tx}^{(0)}=A_x^{(0)}=0}\ , \nonumber \\
\gamma =  \frac{J^{A}}{E^{B}}|_{g_{tx}^{(0)}=A_x^{(0)}=0} = - \frac{i}{\omega} \frac{A_x^{(1)}}{B_x^{(0)}}|_{g_{tx}^{(0)}=A_x^{(0)}=0}\ , \nonumber \\
\beta T =  \frac{Q}{E^{B}}|_{g_{tx}^{(0)}=A_x^{(0)}=0} = \frac{i \delta \rho}{\omega} - \delta \mu \sigma_B -\mu \gamma\, .
\label{betaT}
\end{eqnarray}
Notice the relation
\begin{equation}
\gamma=\sigma_A \frac{J^B}{J^A}|_{g_{tx}^{(0)}=B_x^{(0)}=0}=\sigma_B \frac{J^A}{J^B}|_{g_{tx}^{(0)}=A_x^{(0)}=0}\, ,
\end{equation}
which constitutes a valuable test in the numerical calculations. 

Finally, we find that the (non canonical) thermal conductivity is given by
\begin{equation}\label{kappa}
\kappa\,T=\frac{i}{\omega}[\epsilon + p -2\mu\rho -2\delta\mu\delta\rho] + \sigma_A \mu^2 + \sigma_B
\delta\mu^2 + 2 \gamma \mu \delta\mu\,,
\end{equation}
where the term in the pressure $p=\epsilon/2$ has been added by hand to account for contact terms not directly implemented by the previous computations (see Herzog's review in \cite{reviewsADSCM}). 

The relations we have found between the different conductivities emerge quite naturally from the holographic setup. In the dual field theories they arise from Ward identity constraints on the correlators (see, again, Herzog's review).

Solving numerically the equations (\ref{eqA}), (\ref{eqB}) for the fluctuations of $A_x$ and $B_x$ and using the formulas above, we can calculate all the conductivities ($\sigma_A, \sigma_B, ,\gamma, \kappa, \alpha T, \beta T$) in terms of values of the dual gravity fields.\footnote{An alternative method consists in considering the linear relations in (\ref{matrix}) for different arbitrary choices of the values of the fluctuations at the horizon, in order to obtain enough equations to determine the various conductivities. Solving the system numerically we have checked that this method is stable w.r.t. those choices and it gives the same results as the method described above.}

\subsection{Conductivities in the normal phase}\label{res_normal}
As we have noticed in Section \ref{normal}, following \cite{iqbal}, the normal phase of our model can be seen
as a simplified holographic realization of a ``forced'' ferromagnet. Studying the conductivity matrix in this case is thus quite interesting, since it allows us to make some parallel with known
results in ferromagnetic spintronics.

\begin{figure}[t]
\centering
\includegraphics[width=78mm]{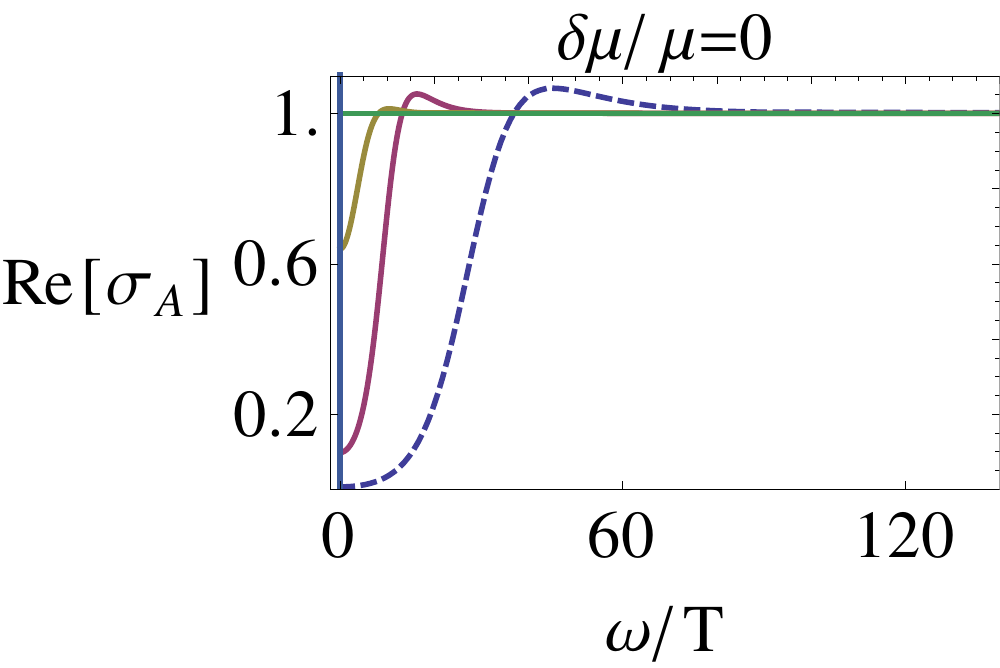} \hspace{0.5cm}
\includegraphics[width=78mm]{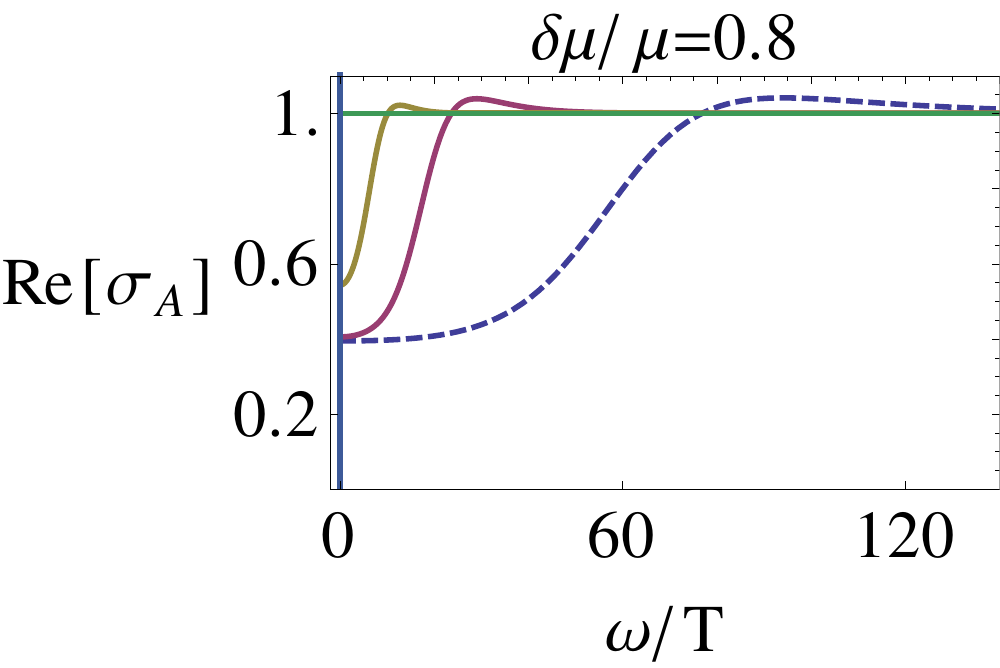}
\caption{The real part of the electric conductivity for $\delta\mu/\mu=0, 0.8$ (left plot, right plot) at $T_c$ (dashed curves) and $T>T_c$ (solid curves).}
\label{ReaNT0}
\end{figure}

In the left plot of Figure \ref{ReaNT0} we present the results for the real part of $\sigma_A$ (the optical electric conductivity)
as a function of the frequency of the external field perturbation for vanishing imbalance, $\delta\mu=0$,
i.e. the case considered in \cite{hhh2}.
The different curves correspond to different temperatures;
the dashed curve is at the critical temperature for the onset of superconductivity.
At very large temperature, the conductivity is basically constant (quantities are normalized such that the constant is precisely equal to 1);
this feature is peculiar of the gravity description we are using.\footnote{It depends on electro-magnetic duality of the four-dimensional
Einstein-Maxwell theory on $AdS_4$ \cite{Herzog:2007ij}.}

As the temperature is decreased, the conductivity is depleted at small frequencies.
In fact, the imaginary part of $\sigma_A$ (not shown) has a pole at $\omega=0$; this is mapped by a Kramers-Kroning relation
to a delta function for $\rm{Re}[\sigma_A]$ at the same point (the solid line at $\omega/T=0$ in the figure).
The delta function at zero frequency is due to the system translation invariance which, in charged media,
causes an overall uniform acceleration and so an infinite DC conductivity.\footnote{This infinity is of course expected to transform into the standard Drude peak upon inclusion of impurities breaking translation invariance. Note that we are working with the fully backreacted solution, so there is no dissipation as it happens in the probe limit, and as a result translational invariance is preserved.}
Since the area under the curves must be constant at different temperatures due to a Ferrell-Glover-Tinkham sum rule, the development of a delta function at $\omega=0$ is compensated by a depletion of the conductivity at small frequencies \cite{hhh2}.

In the right plot of Figure \ref{ReaNT0} we present the results for the real part of $\sigma_A$ for non-vanishing imbalance, $\delta\mu/\mu=0.8$.
The general trend is clearly the same as in the balanced case.
Nevertheless, the depth of the depletion at small frequencies is reduced with respect to the $\delta\mu/\mu=0$ case
(remember that $T_c$ is reduced too): the magnitude of the delta function is reduced, i.e. the DC conductivity is decreased by $\delta\mu$.

\begin{figure}[t]
\centering
\includegraphics[width=78mm]{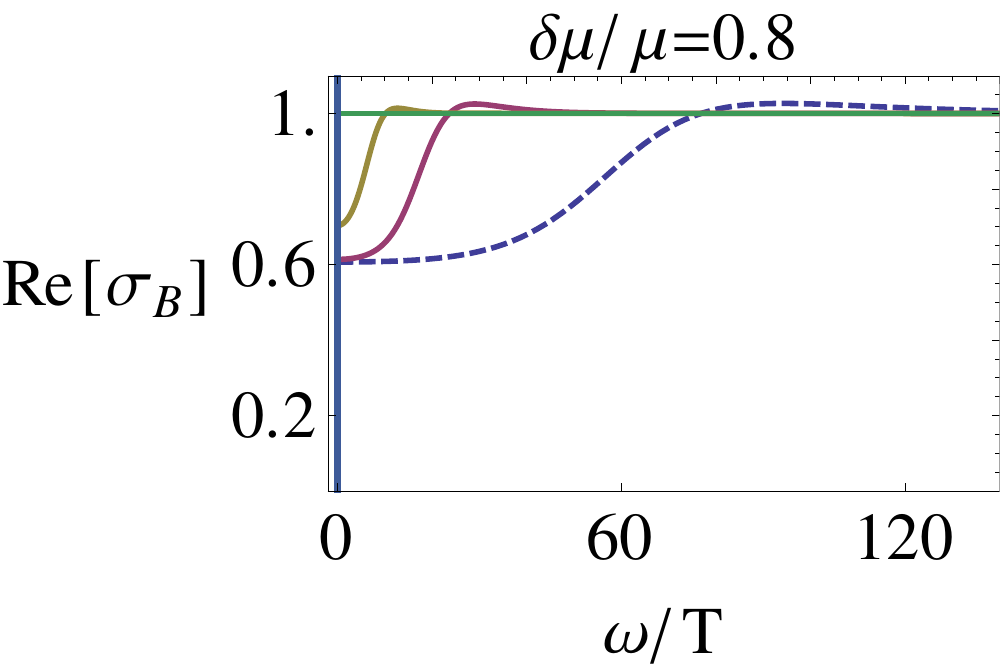} \hspace{0.5cm}
\includegraphics[width=78mm]{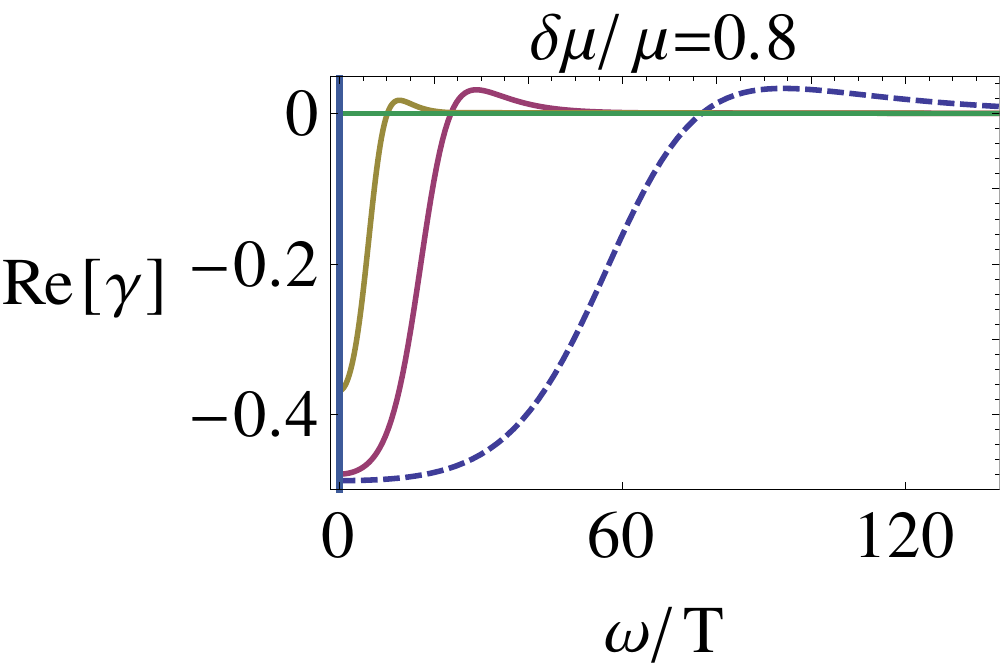}
\caption{The real part of the ``spin-spin conductivity'' $\sigma_B$ (left) and of the ``spin conductivity'' $\gamma$ (right) for $\delta\mu/\mu=0.8$ at $T_c$ (dashed curves) and $T>T_c$ (solid curves).}
\label{RebNT}
\end{figure}

Concerning the real part of $\sigma_B$  (the ``optical spin-spin conductivity''), it is exactly constant in absence of imbalance, $\delta\mu=0$: there is no ``net spin'' in the system, so the conductivity is featureless.\footnote{The gravity vector $B_a$ fluctuates on a black hole background charged under the other $U(1)_A$.}
The unbalanced case, $\delta\mu/\mu=0.8$, is reported in the left plot of Figure \ref{RebNT}.
The similarity with the plot of $\sigma_A$, including the infinite DC conductivity, is transparent.
In fact, it is clear (for example from equations (\ref{eqA}), (\ref{eqB})) that in the normal phase
(that is, at zero $\psi$) the system enjoys the symmetry
\begin{equation}\label{rela}
\sigma_A (\mu, \delta\mu, \omega/T) = \sigma_B (\delta\mu, \mu, \omega/T)\, ,
\end{equation}
which we also verified numerically with ${\cal O}(10^{-3})$ accuracy (at least).
In particular,
\begin{equation}\label{relazione}
\sigma_A  = \sigma_B \qquad {\rm for} \qquad \mu=\delta\mu\, .
\end{equation}
This is the same as stating that for perfectly polarized electrons (say all spins ``up''), the electric conductivity $\sigma_{cc}$ ($\sigma_A$) equals the ``spin-spin'' one $\sigma_{ss}$ ($\sigma_B$).
In this case, then, we recover the zero-momentum result in \cite{shibataealtro}.

Actually, formula (\ref{rela}) is just a part of a set of more general relations among the various conductivities in the normal phase.
In fact, it turns out that all the conductivities can be given once a single frequency dependent function $f=f(\omega/T,\delta\mu/\mu)$ (a ``mobility function'' for the charge/spin carriers) is known.  
Considering formulas (\ref{alphaT}), (\ref{betaT}), (\ref{kappa}), the conductivity matrix $\hat \sigma$ in the normal phase turns out to be\footnote{We have left $\kappa T$ implicit just to save space: its expression, in terms of the other conductivities and of the thermodynamical parameters in the normal phase (as given in Section \ref{normal}) can be immediately deduced from eq. (\ref{kappa}).}
\begin{eqnarray}\label{superrelazione}
\hat\sigma = \begin{pmatrix}  \sigma_A & \alpha T & \gamma \\ \alpha T &  \kappa T & \beta T \\ \gamma & \beta T & \sigma_B \end{pmatrix}=
\end{eqnarray}
\begin{eqnarray}
\begin{pmatrix} f \rho^2+1 & \frac{i \rho}{\omega}-\mu(f \rho^2+1)-\delta\mu f \rho\,\delta\rho & f \rho\ \delta\rho \\ \frac{i \rho}{\omega}-\mu(f \rho^2+1)-\delta\mu f \rho\,\delta\rho &  \kappa T &  \frac{i \delta \rho}{\omega}-\delta\mu(f \delta\rho^2+1)-\mu f \rho\,\delta\rho  \\ f \rho\ \delta \rho & \frac{i \delta \rho}{\omega}-\delta\mu(f \delta\rho^2+1)-\mu f \rho\,\delta\rho  & f \delta\rho^2+1 \end{pmatrix} \, .\nonumber
\end{eqnarray}
These relations concern both the real and the imaginary parts of the conductivities.
The matrix (\ref{superrelazione}) reproduces in many instances the expectations in \cite{vanson}, for example despite their complicated explicit expressions, we have $\beta T = (\delta\rho/\rho) \alpha T$ (since
in the normal phase $\rho\delta\mu=\mu\delta\rho$, see Section \ref{normal}). The diagonal conductivities are related quadratically to the corresponding
charge densities because the more the carriers are coupled the more they feel the
external field, and, in addition, the bigger the charge density the stronger the transport.
Again, the form (\ref{superrelazione}) has been verified numerically with at least ${\cal O}(10^{-3})$ accuracy.

The right plot of Figure \ref{RebNT} reports the results for $\gamma$, the mixed conductivity (the ``spin conductivity'' $\sigma_{sc}=\sigma_{cs}$), at $\delta\mu/\mu=0.8$.
At $\delta\mu=0$ this conductivity is identically zero: in absence of ``net spin'', an external electric field does not cause the transport of ``spin'' (and the other way around).
Instead, as $\delta\mu/\mu\neq 0$, we have the typical phenomenon at the roots of spintronics: an external electric field causes the transport of ``spin'' (and the other way around) with a non-trivial conductivity $\gamma$.

Note that in the $\mu=\delta\mu$ case (which from formulas (\ref{phiN}), (\ref{vN}) is equivalent to $\rho=\delta\rho$),
(\ref{superrelazione}) implies $\gamma=\sigma_A-1=\sigma_B-1$: for perfectly polarized electrons the ``spin conductivity''
$\gamma$ is equal to the electric conductivity $\sigma_A$, except for the constant contribution at large $\omega/T$ (the ``1'' in formula (\ref{superrelazione}) with our normalizations of the constant conductivity).
The latter contribution marks the difference with the quasi-free electron case recently studied in \cite{shibataealtro}, where $\gamma=\sigma_B$ at $\mu=\delta\mu$.
The discrepancy is simply due to the fact that, differently from the quasi-free electron case,
in the system at hand we have conduction even in the absence of net charge/spin density.

\begin{figure}[t]
\centering
\includegraphics[width=78mm]{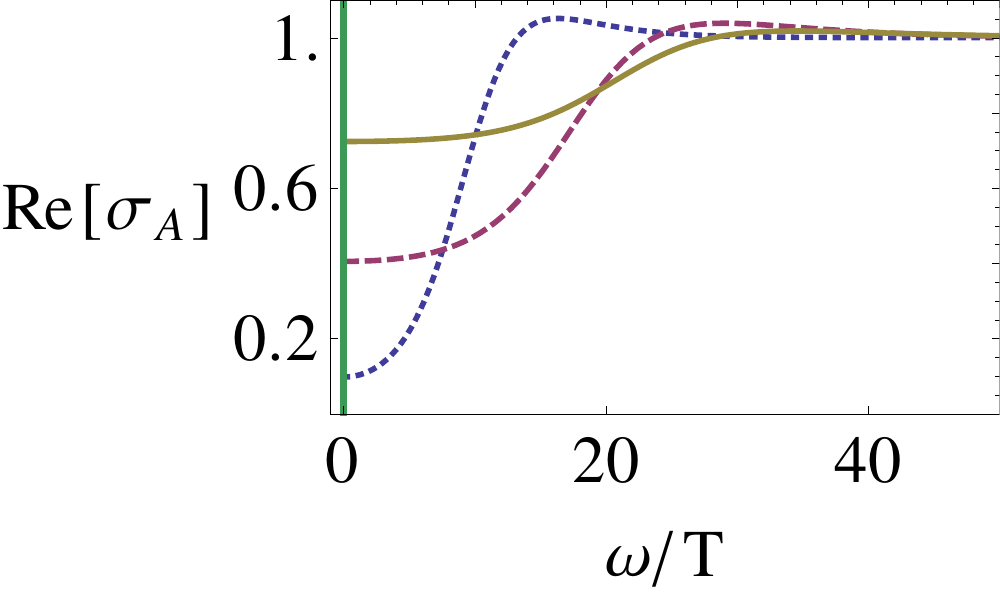} \hspace{0.5cm}
\includegraphics[width=78mm]{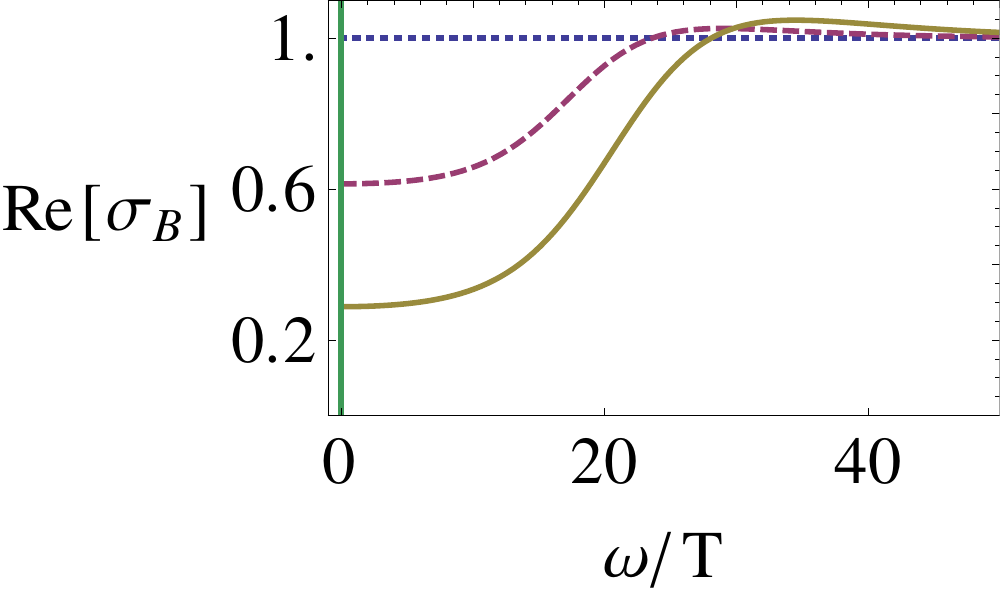}
\caption{The real part of the electric conductivity $\sigma_A$ (left) and of the ``spin-spin conductivity'' $\sigma_B$ (right) for $\delta\mu/\mu=0, 0.8, 1.6$ (dotted, dashed and solid lines respectively) at fixed temperature.}
\label{ReaN}
\end{figure}

In Figure \ref{ReaN} we present the real part of $\sigma_A$ (left) and $\sigma_B$ (right) at fixed temperature for different values of $\delta\mu/\mu=0,0.8,1.6$ (dotted, dashed and solid lines respectively).\footnote{We consider a generic setup where $\delta\mu/\mu$ is not constrained to be smaller or equal to one. In QCD with up-down quark condensates, for example, $\delta\mu$ is given by the isospin chemical potential, while $\mu$ is the baryonic one, the two being in principle independent.}
Note the opposite behavior, dictated by formula (\ref{rela}), of the conductivities $\sigma_A, \sigma_B$ with increasing $\delta\mu/\mu$, which determines the increase of the  $\sigma_B$ DC conductivity with $\delta\mu/\mu$.
This behavior, which has obvious physical origin in our system, is present in other contexts as well (see e.g. \cite{Bellazzini:2008mn}), where it is usually interpreted as a separation of the dynamics of charge and spin degrees of freedom.

\begin{figure}[t]
\centering
\includegraphics[width=78mm]{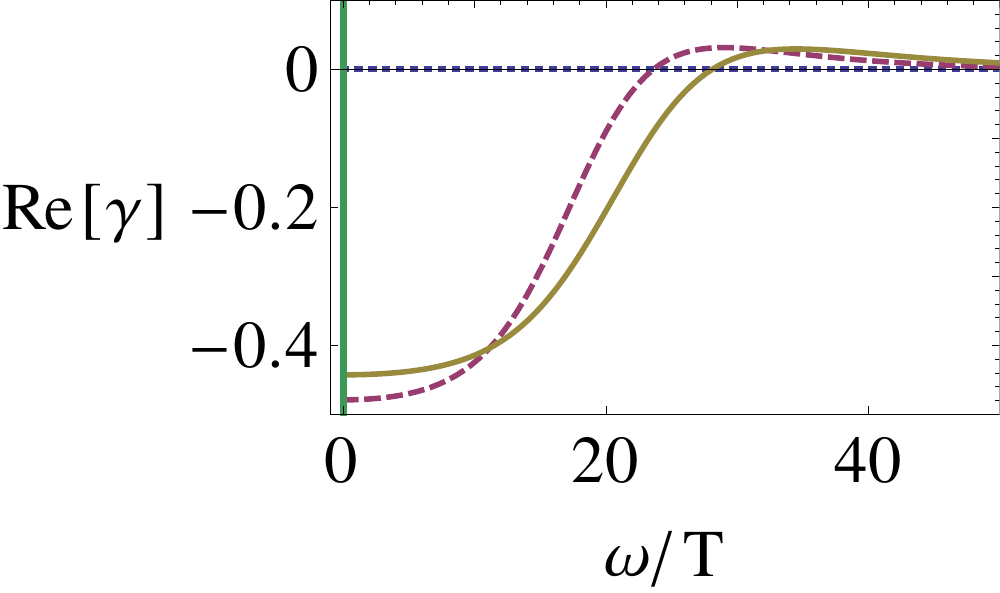} \hspace{0.5cm}
\includegraphics[width=78mm]{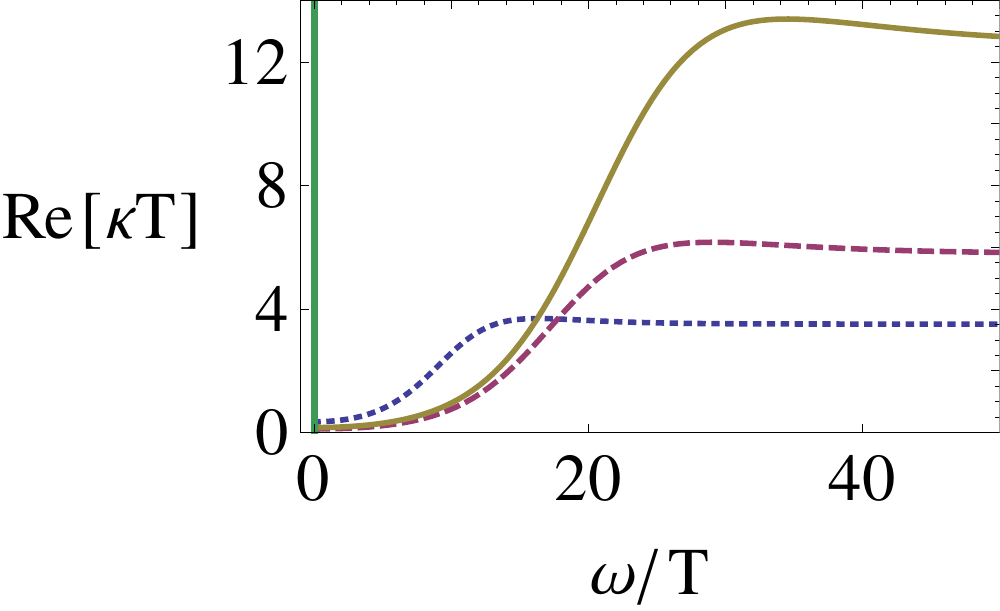}
\caption{The real part of the ``spin conductivity'' $\gamma$ (left) and of the thermal conductivity $\kappa$ (right) for $\delta\mu/\mu=0, 0.8, 1.6$ (dotted, dashed and solid lines respectively) at fixed temperature.}
\label{RegN}
\end{figure}

In Figure \ref{RegN} we present the real part of $\gamma$ (left) and of the thermal conductivity $\kappa$ (right) at fixed temperature for different values of $\delta\mu/\mu=0,0.8,1.6$ (dotted, dashed and solid lines respectively).
Notice the non-monotonic behavior with $\delta\mu/\mu$ of both Re$[\gamma]$ and Re$[\kappa]$ in the small frequency region. Notice moreover that there is a delta function (whose coefficient is enhanced by $\delta\mu$) in the DC thermal conductivity due to momentum conservation (translation invariance) which forbids the relaxation of the heat current \cite{hhh2}. This is reflected in a pole in Im$[\kappa]$ at $\omega=0$.

\begin{figure}[t]
\centering
\includegraphics[width=78mm]{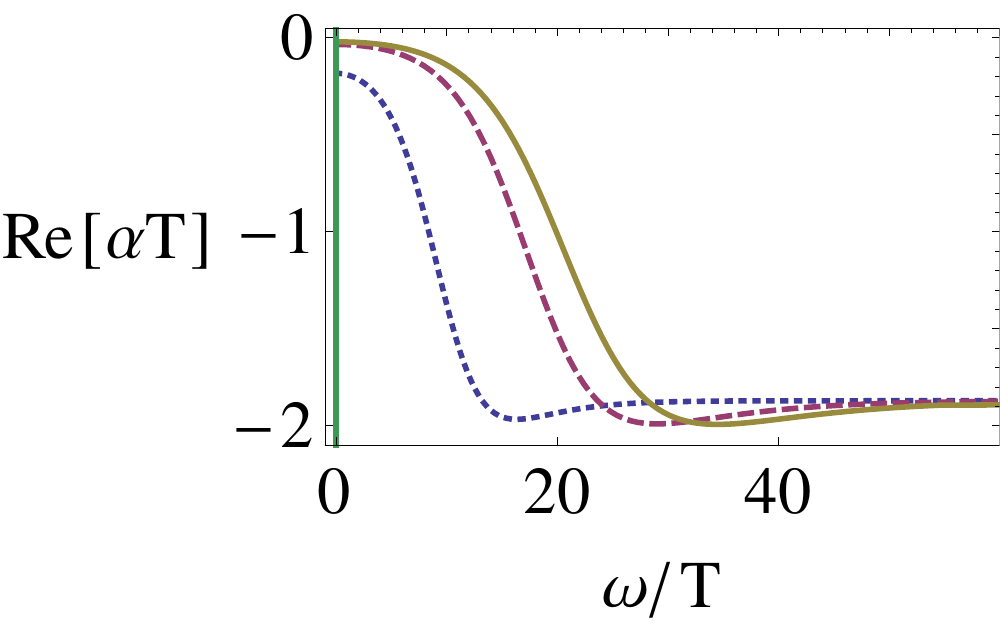} \hspace{0.5cm}
\includegraphics[width=78mm]{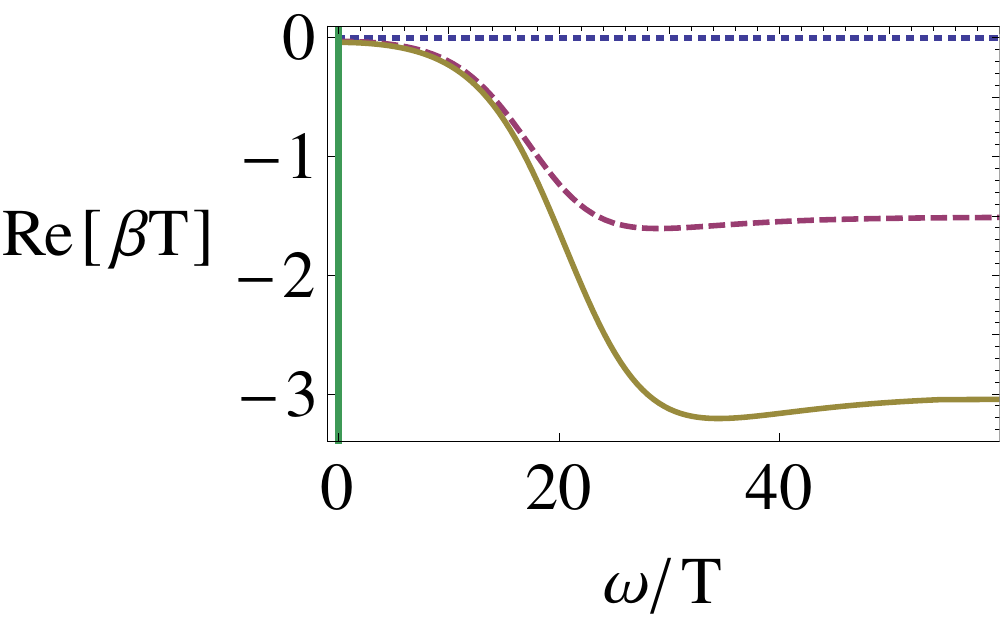}
\caption{The real part of the thermo-electric conductivity $\alpha T$ (left) and of the ``spin-electric'' conductivity $\beta T$ (right) for $\delta\mu/\mu=0, 0.8, 1.6$ (dotted, dashed and solid lines respectively) at fixed temperature.}
\label{RealphaTN}
\end{figure}

Finally, in Figure \ref{RealphaTN} we report the real part of the thermo-electric conductivity $\alpha T$ (left) and of the ``spin-electric'' conductivity $\beta T$ (right) at fixed temperature for different values of $\delta\mu/\mu=0,0.8,1.6$ (dotted, dashed and solid lines respectively).

\subsection{Conductivities in the superconducting phase}\label{res_sucond}

\begin{figure}[t]
\centering
\includegraphics[width=78mm]{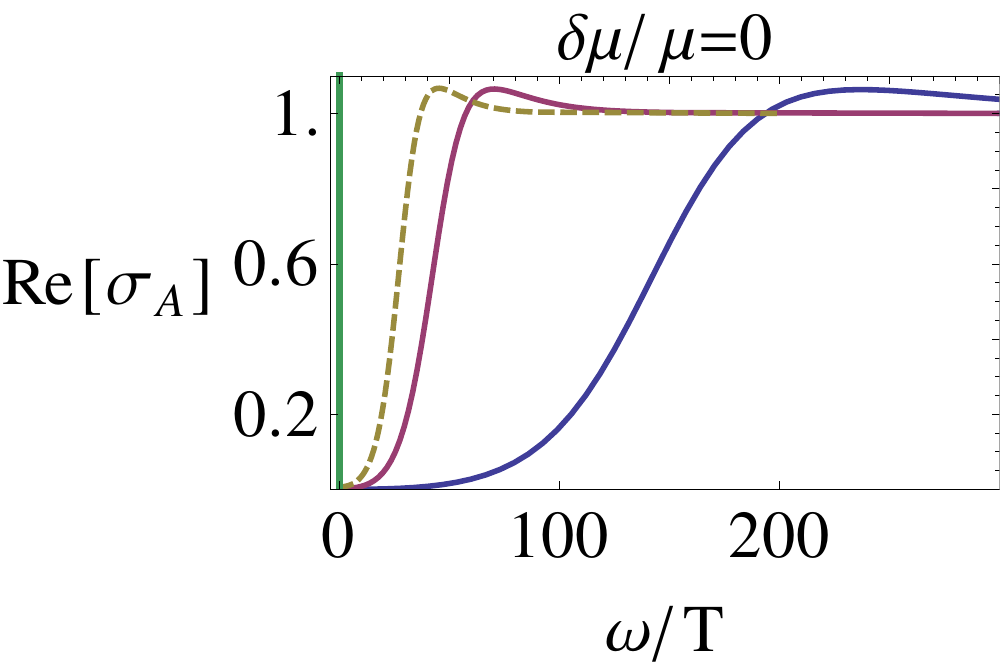} \hspace{0.5cm}
\includegraphics[width=78mm]{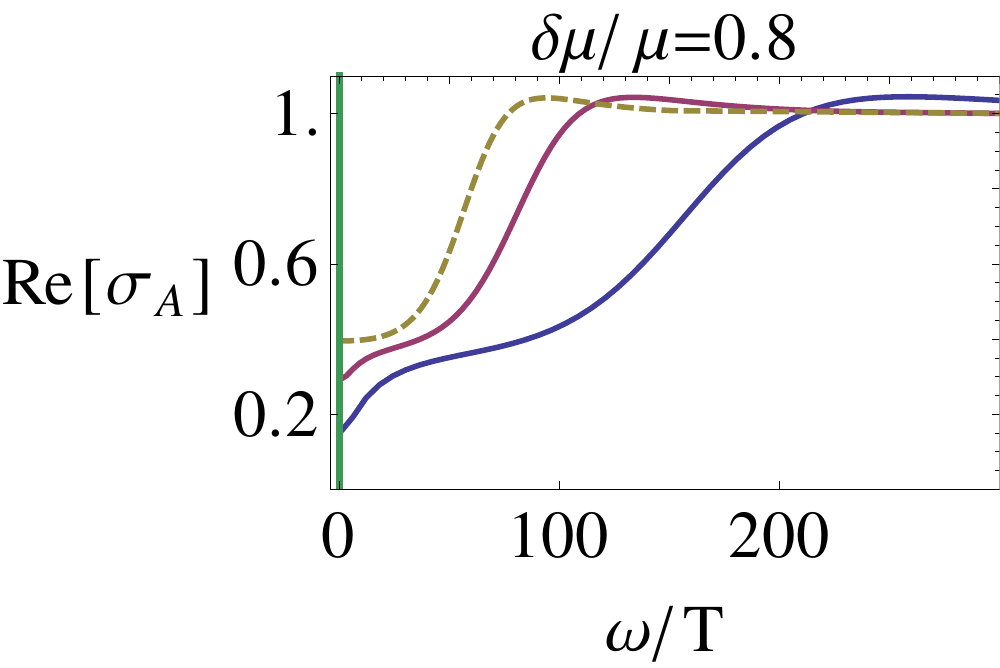}
\caption{The real part of the electric conductivity for $\delta\mu/\mu=0, 0.8$ (left plot, right plot) at $T_c$ (dashed curves) and $T<T_c$ (solid curves).}
\label{ReaT0}
\end{figure}

\begin{figure}[t]
\centering
\includegraphics[width=78mm]{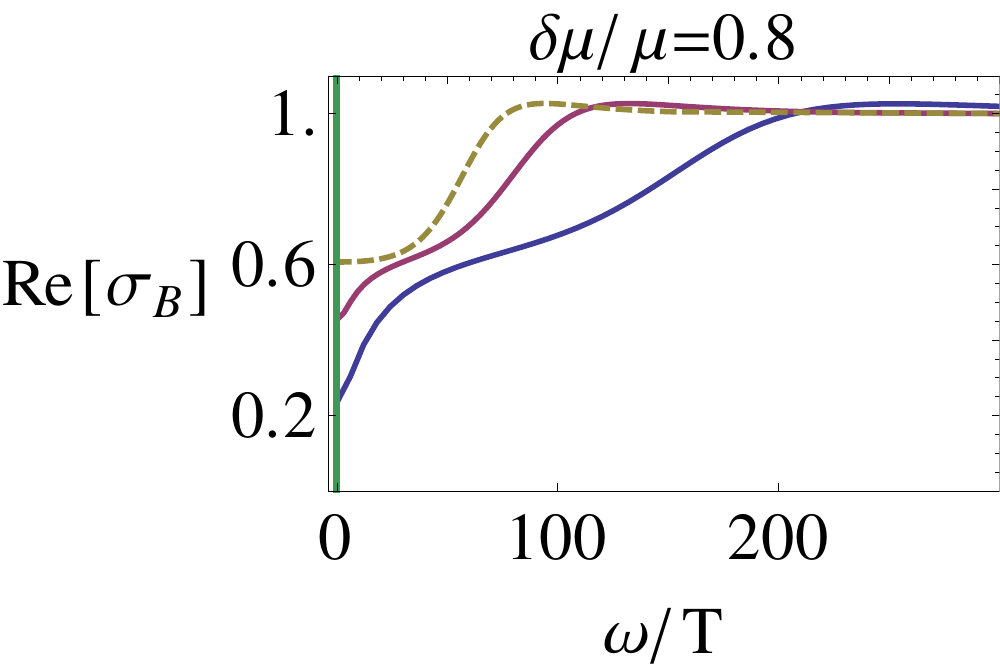} \hspace{0.5cm}
\includegraphics[width=78mm]{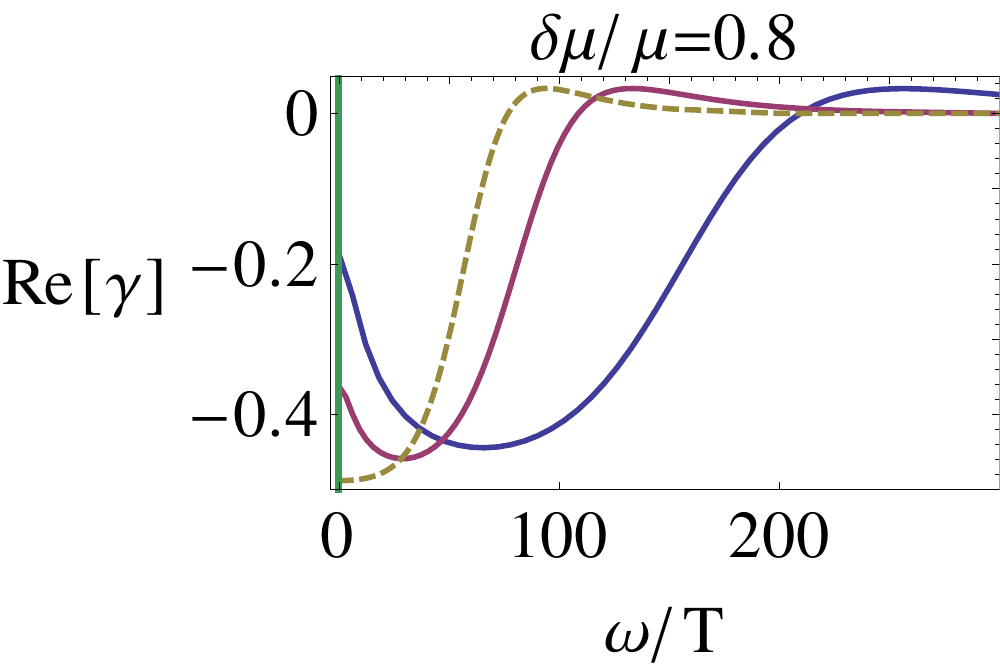}
\caption{The real part of the ``spin-spin conductivity'' $\sigma_B$ (left) and of the ``spin conductivity'' $\gamma$ (right) for $\delta\mu/\mu=0.8$ at $T_c$ (dashed curves) and $T<T_c$ (solid curves).}
\label{RegT}
\end{figure}

The thermal behavior of the conductivities in the superconducting phase is similar to the one in the normal phase and it is shown in Figures \ref{ReaT0}, \ref{RegT}.

\begin{figure}[t]
\centering
\includegraphics[width=78mm]{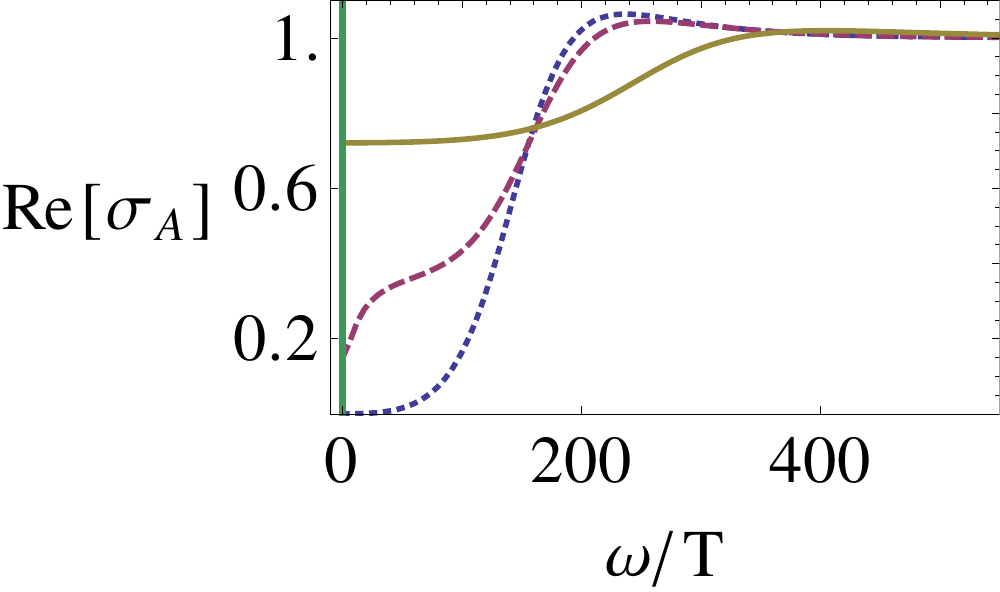} \hspace{0.5cm}
\includegraphics[width=78mm]{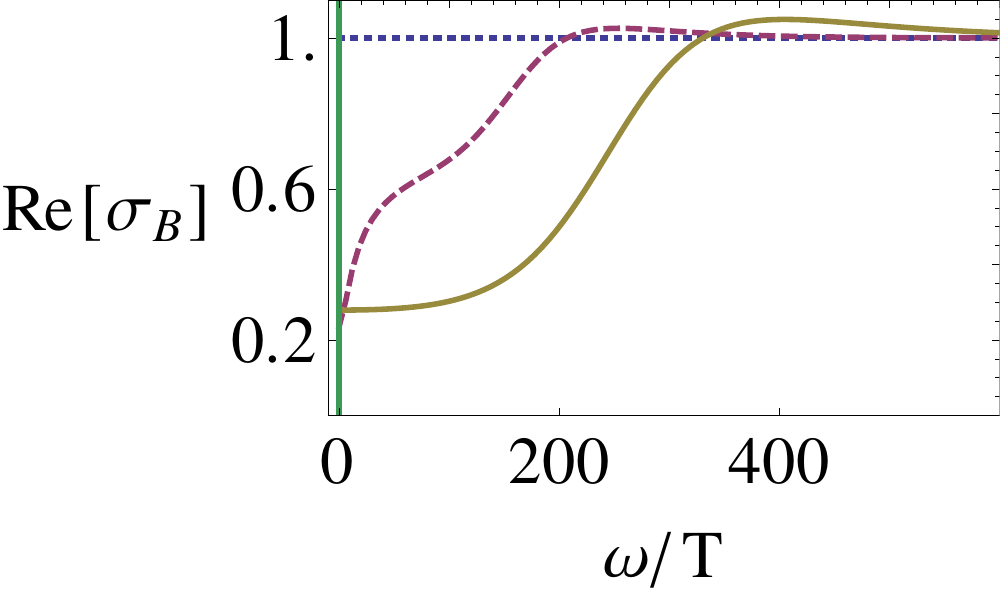}
\caption{The real part of the electric conductivity $\sigma_A$ (left) and of the ``spin-spin conductivity'' $\sigma_B$ (right) for $\delta\mu/\mu=0, 0.8, 1.6$ (dotted, dashed and solid lines respectively) at fixed temperature below $T_c$.}
\label{Rea}
\end{figure}

\begin{figure}[t]
\centering
\includegraphics[width=78mm]{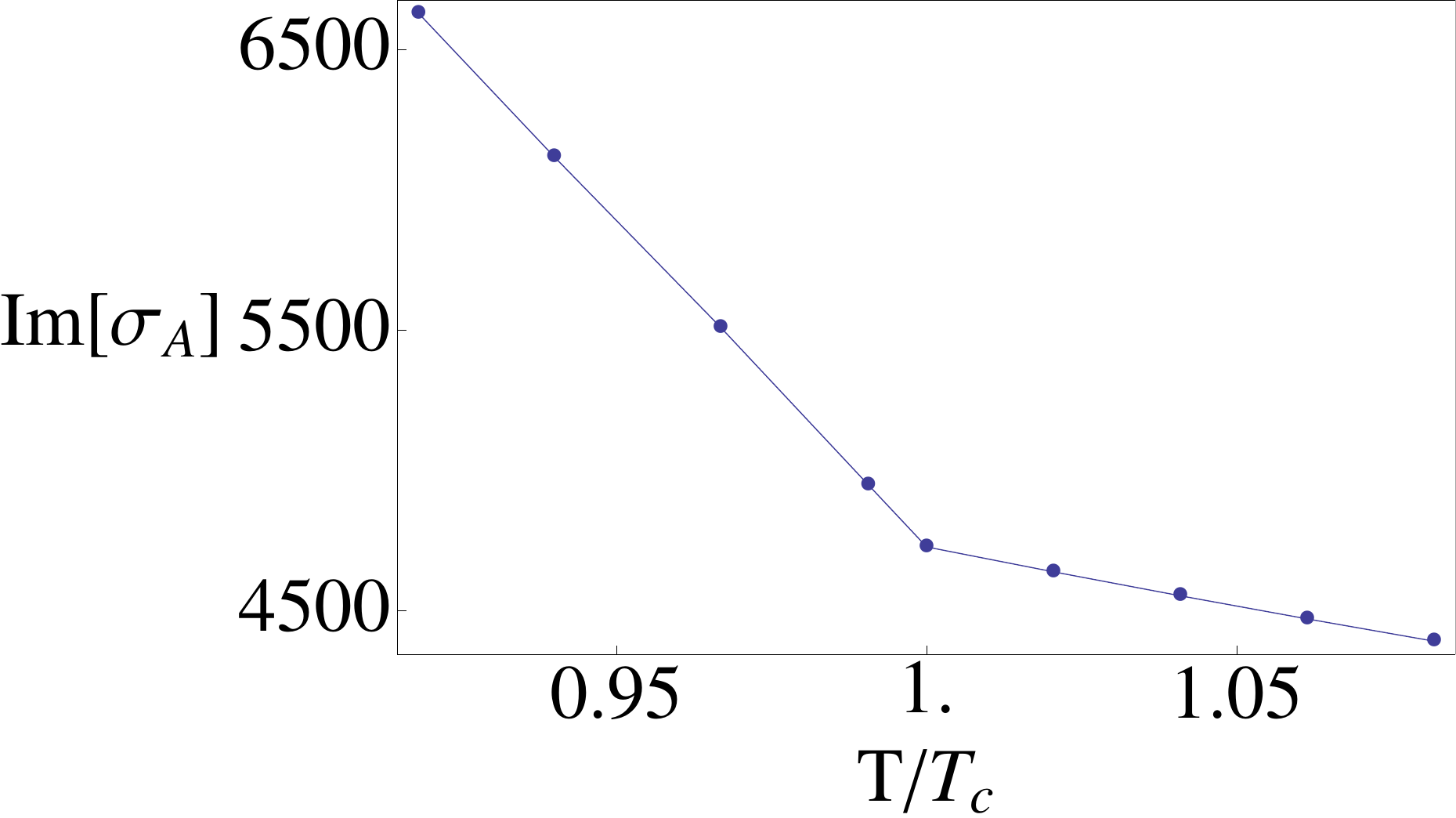} \hspace{0.5cm}
\includegraphics[width=78mm]{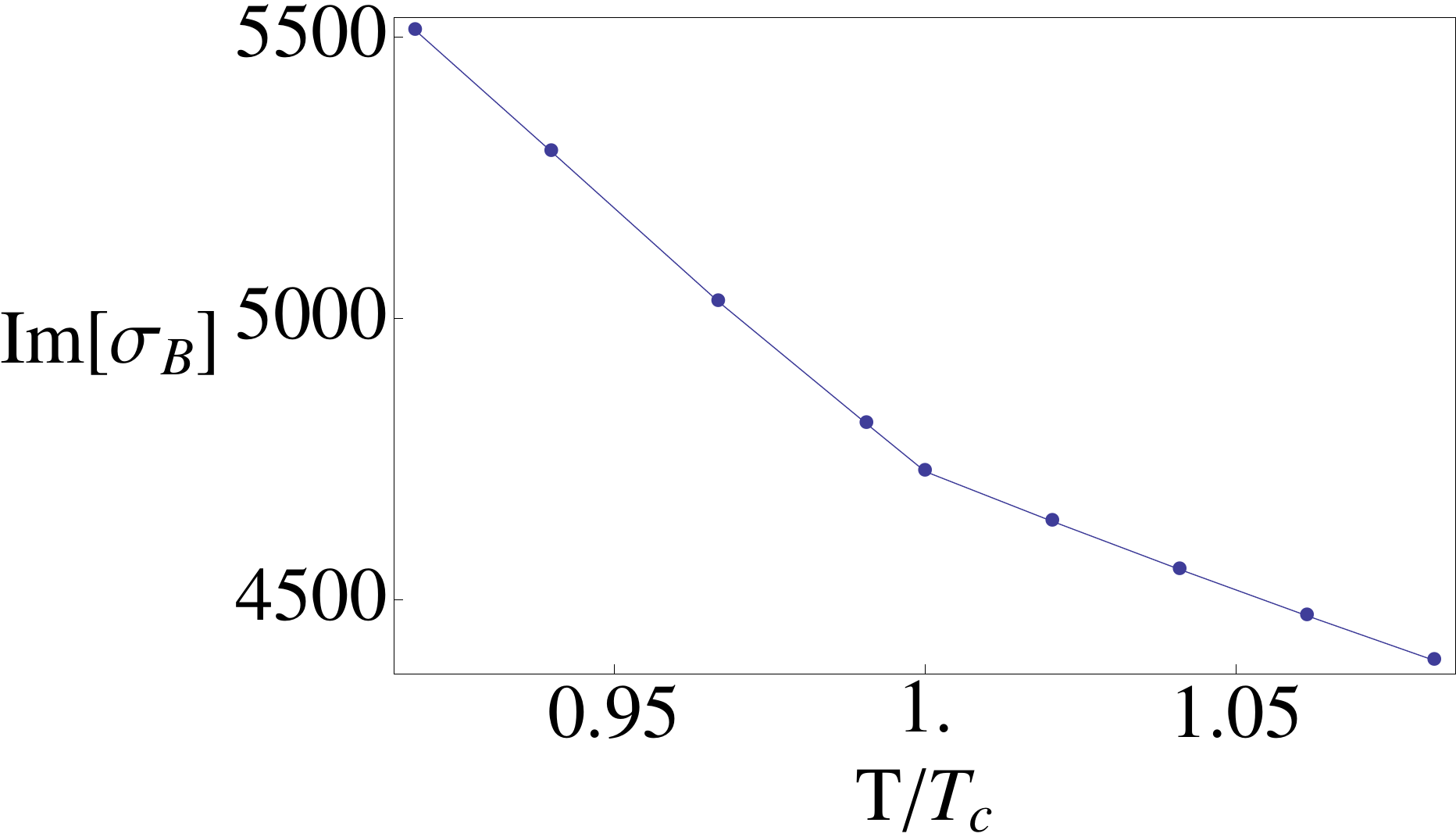}
\caption{The discontinuity at $T_c$ in the imaginary part of the electric conductivity $\sigma_A$
(left) and of the ``spin-spin conductivity'' $\sigma_B$ (right), signaling a discontinuity
in the magnitude of the delta function at $\omega= 0$ in the respective DC
conductivities.}
\label{jump}
\end{figure}

\begin{figure}[t]
\centering
\includegraphics[width=78mm]{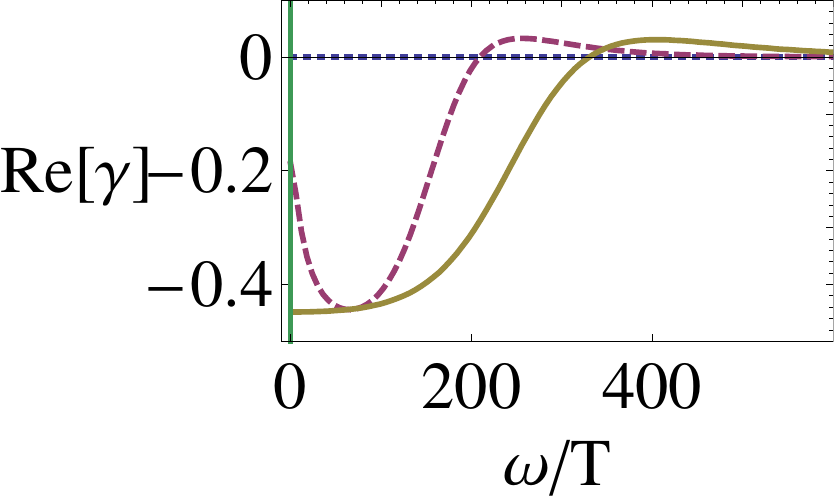} \hspace{0.5cm}
\includegraphics[width=78mm]{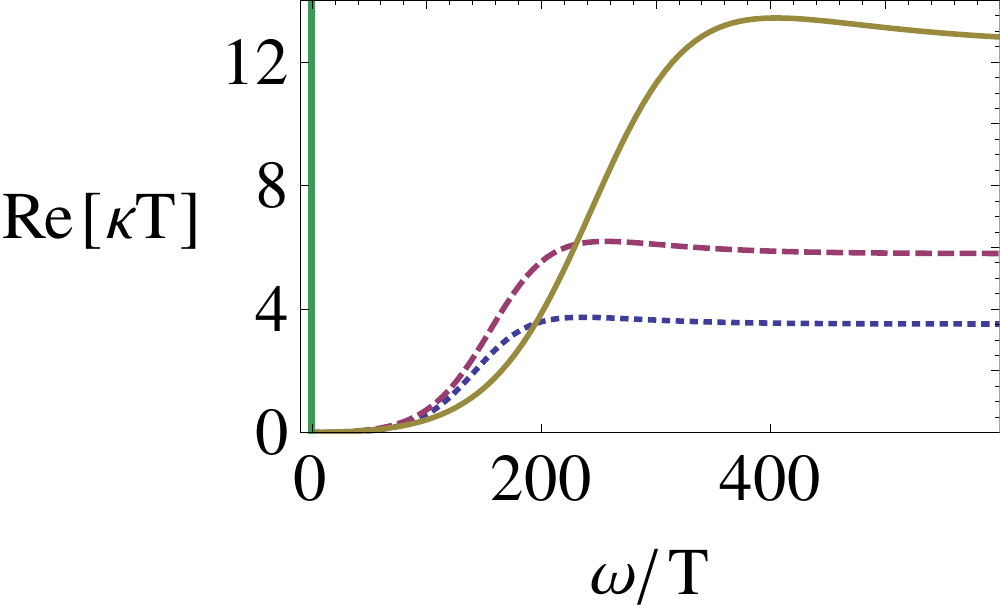}
\caption{The real part of the ``spin conductivity'' $\gamma$ (left) and of the thermal conductivity $\kappa$ (right) for $\delta\mu/\mu=0, 0.8, 1.6$ (dotted, dashed and solid lines respectively) at fixed temperature below $T_c$.}
\label{Reg}
\end{figure}

In Figures \ref{Rea}, \ref{Reg}, we present the results for $\sigma_A, \sigma_B, \gamma$ in the superconducting phase for $\delta\mu/\mu=0, 0.8, 1.6$ (dotted, dashed and solid lines respectively) at constant temperature below $T_c$.
The dotted line in Figure \ref{Rea} corresponds to the balanced case ($\delta\mu=0$) of \cite{hhh2}.
The optical electric conductivity at small temperature presents a pseudo-gap\footnote{Since the system is really a superfluid, the spectrum is ungapped, containing the Goldstone boson of the breaking of $U(1)_A$. This forbids the presence of a hard gap at $T=0$ \cite{horrob}. Nevertheless, the conductivity is extremely small (almost exponentially) at small frequencies for sufficiently low temperatures, hence we speak about a ``pseudo-gap''.} at small frequencies, while it relaxes to the normal phase value at large $\omega$.
The corresponding imaginary part of the conductivity (not shown) has a pole at $\omega=0$, which translates in a delta function at the same point in $\rm{Re}[\sigma_A]$  (the solid line at $\omega/T=0$ in the figure) due to a Kramers-Kroning relation.
A part of this infinite DC conductivity is due to translation invariance and it is present also in the normal phase, as described in Section \ref{res_normal}.
Nevertheless, in Figure \ref{jump} (left plot) it is shown that the imaginary part of
the conductivity has discontinuous derivative at $T_c$, and so the coefficient of the delta function
has a jump across the phase transition.
This means that a part of the delta function corresponds to genuine superconductivity and is due to the spontaneous breaking of $U(1)_A$ via the condensation of the charged operator dual to $\psi$.

As $\delta\mu/\mu$ is turned on and increased, the pseudo-gap in the real part of the electric conductivity  $\rm{Re}[\sigma_A]$
is reduced and eventually lost (dashed and solid lines in Figure \ref{Rea}).
This behavior has to be expected, since for large $\delta\mu/\mu$ the system at any fixed temperature will pass, with a second order, continuous transition, to the normal phase; accordingly, the conductivity has a very similar behavior.
Precisely the same pattern is seen in the optical conductivities of some iron-based superconductors with increasing doping fraction (e.g. \cite{Nakajima}\footnote{While in Fe-based superconductors the doping dependence does not directly translate into a chemical composition dependence as in cuprates, we chose to mention such an example due to the fact that Fe superconductors are believed to be s-wave (more precisely, $s_{+-}$), as the ones considered in this paper.}).

Inversely, from the right plot in Figure \ref{Rea} we see that the real part of the ``spin-spin conductivity'' $\rm{Re}[\sigma_B]$ is more and more depleted at small frequencies as $\delta\mu/\mu$ is increased.
Also in this case it can be shown (right plot of Figure \ref{jump}), that the imaginary part
of the conductivity has discontinuous derivative at $T_c$; the coefficient of the delta
function at $\omega = 0$ presents therefore a jump across the phase transition temperature.\footnote{The existence of this jump can be also
verified, following section 4.2 of \cite{hhh2}, by considering the analytic form of the small frequency regime of the imaginary part of the
conductivity slightly above the critical temperature, and checking numerically that the coefficient of the pole has a discontinuity when
going slightly below $T_c$. The size of this discontinuity is of the same order of the one for $\sigma_A$.} That is, even if the condensing operator is uncharged with respect to $U(1)_B$, due to the mixing with the $U(1)_A$
current in the superconducting phase, the $\sigma_B$ DC conductivity is enhanced. This strong coupling behavior is in stark contrast with the usual weak coupling picture,
where the conductivity $\sigma_B$ is reduced by the decrease of the quasi-particle
population due to the formation of the gap \cite{Taka}.

Figure \ref{Reg} (left) shows that the ``spin conductivity'' $\gamma$ behaves qualitatively as $\sigma_B$.
For $\delta\mu=0$ it is constant and vanishing: in absence of ``net spin'', an external electric field does not cause the transport of ``spin'' (and the other way around).
Instead, a depletion at small frequencies opens as $\delta\mu/\mu$ is increased and the DC conductivity becomes infinite.

The ``current polarization'' $P=$Re$[J^B]/$Re$[J^A]$, measuring the
difference of the two DC spin-polarized currents in absence of the ``spin
motive force'', is exactly equal to Re$[\gamma]/$Re$[\sigma_A]$. In
particular, it does not depend on the applied external field $E^A$. In
the normal phase, the relations summarized in (\ref{superrelazione})
imply that $P=\delta\rho/\rho$ exactly. In the superconducting phase, it
can be checked\footnote{Since the DC conductivities are infinite in the superconducting
phase, we estimate the ratio Re$[\gamma]/$Re$[\sigma_A]$ by evaluating the corresponding ratio of
imaginary parts at very small $\omega/T$, relying on the Kramers-Kronig relation for the
result to translate in the respective behavior of the real parts.} that the qualitative
behavior of
$P(\delta\rho/\rho)$ is still increasing, but in a non-linear way (the
power dependence being larger than one).
For examples of current polarization behaviors in BCS systems, see e.g. \cite{giazotto}.
The ``optical current polarization'', instead, can be shown to be non-monotonic with
$\delta\rho/\rho$. 

In the presence of the condensate, the symmetries encountered in the normal phase and encoded in formulas (\ref{rela}), (\ref{superrelazione}) are explicitly broken (relations (\ref{alphaT}), (\ref{betaT}) and (\ref{kappa}) are preserved).
In particular,  (\ref{superrelazione}) should be modified to account for the vev of the charged operator, which heavily influences the small frequency behavior of the conductivities.
Unfortunately, we did not find any simple modification that accounts for the numerical results.

At this point, a comment is in order.
It is clear that for non-extremal values of $\delta\mu/\mu$, exemplified by the dashed lines in the figures of this section,
the conductivities have a non-trivial (even non-monotonic) behavior for small frequencies.\footnote{We think that this behavior is not
due to numerical effects because we checked it with two independent codes and two different methods for the calculations.}
This is reasonable.
In fact, the presence of the new energy scale provided by the imbalance can modify, with respect to the balanced case, the behavior of the system at frequencies related to $\delta\mu$.
We know from \cite{Herzog:2007ij} that the behavior of the balanced system at zero charge density has just one regime at zero momentum: the hydrodynamic and the high frequency regimes coincide and the conductivity is constant.
On the other hand, the condensate in the charged system changes drastically the behavior of the electric conductivity in the small frequency regime, which is now dominated by the superfluid Goldstone mode and the pseudo-gap \cite{hhh1, horrob}.
It is thus natural to expect a non-trivial influence of the extra IR scale $\delta\mu$ on the transport properties of the theory.
In particular, from the plots of this section it seems that the DC electric conductivity is further enhanced by $\delta\mu$ w.r.t. the extrapolated behavior of the normal phase, possibly due to a second light mode.

\begin{figure}[t]
\centering
\includegraphics[width=78mm]{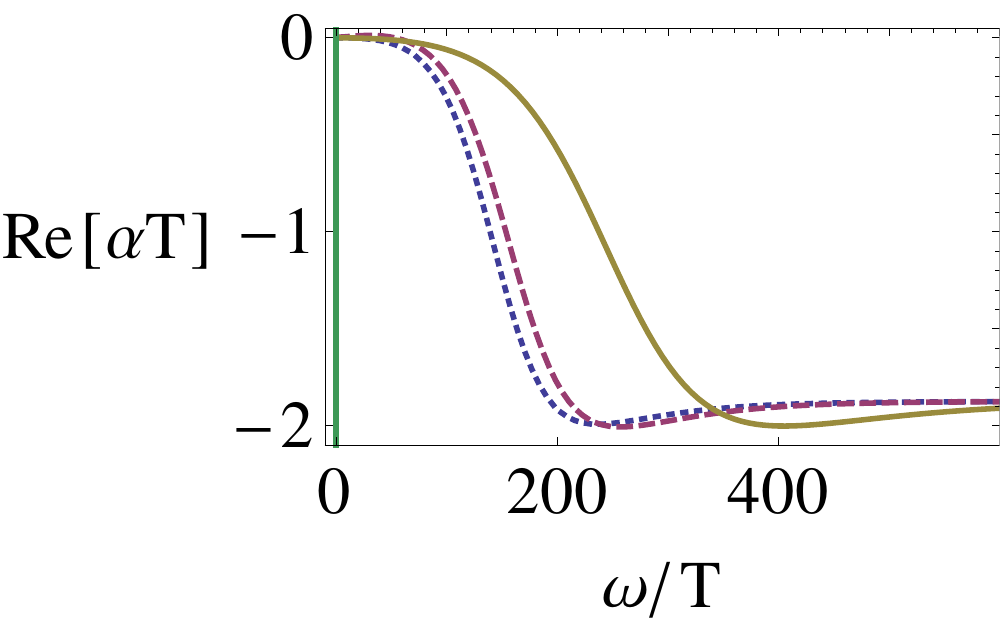} \hspace{0.5cm}
\includegraphics[width=78mm]{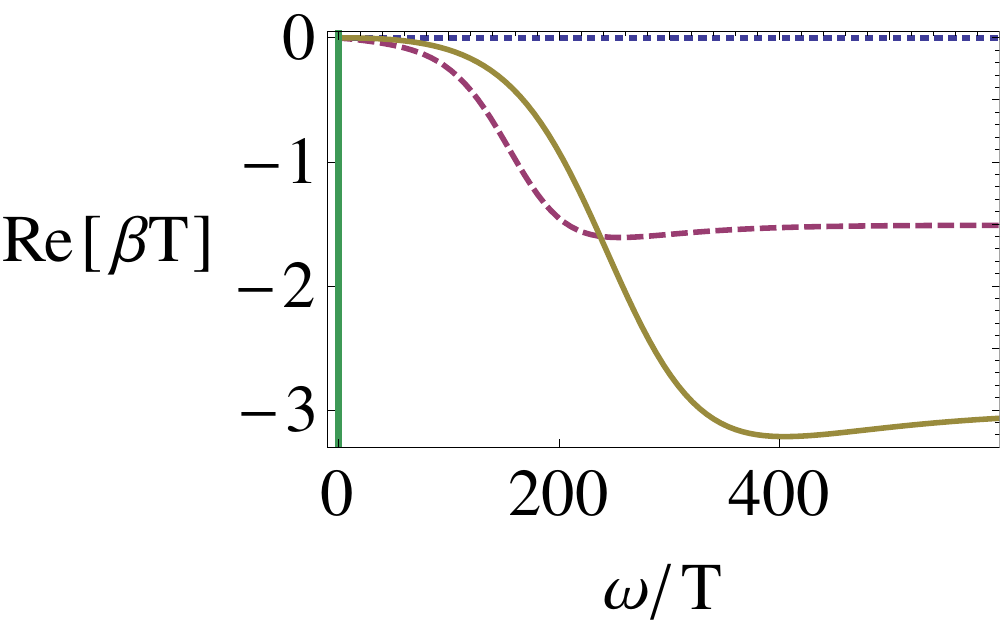}
\caption{The real part of the thermo-electric and ``thermo-spin'' conductivities (left plot, right plot) for $\delta\mu/\mu=0, 0.8, 1.6$ (dotted, dashed and solid lines respectively) at fixed temperature below $T_c$.}
\label{ImKfin}
\end{figure}

The thermo-electric, ``spin-electric'' and thermal conductivity follow from the plots above and formulas (\ref{alphaT}), (\ref{betaT}), (\ref{kappa}).
The corresponding plots are reported in Figure  \ref{Reg} (right) and \ref{ImKfin}.

\begin{figure}[t]
\centering
\includegraphics[width=110mm]{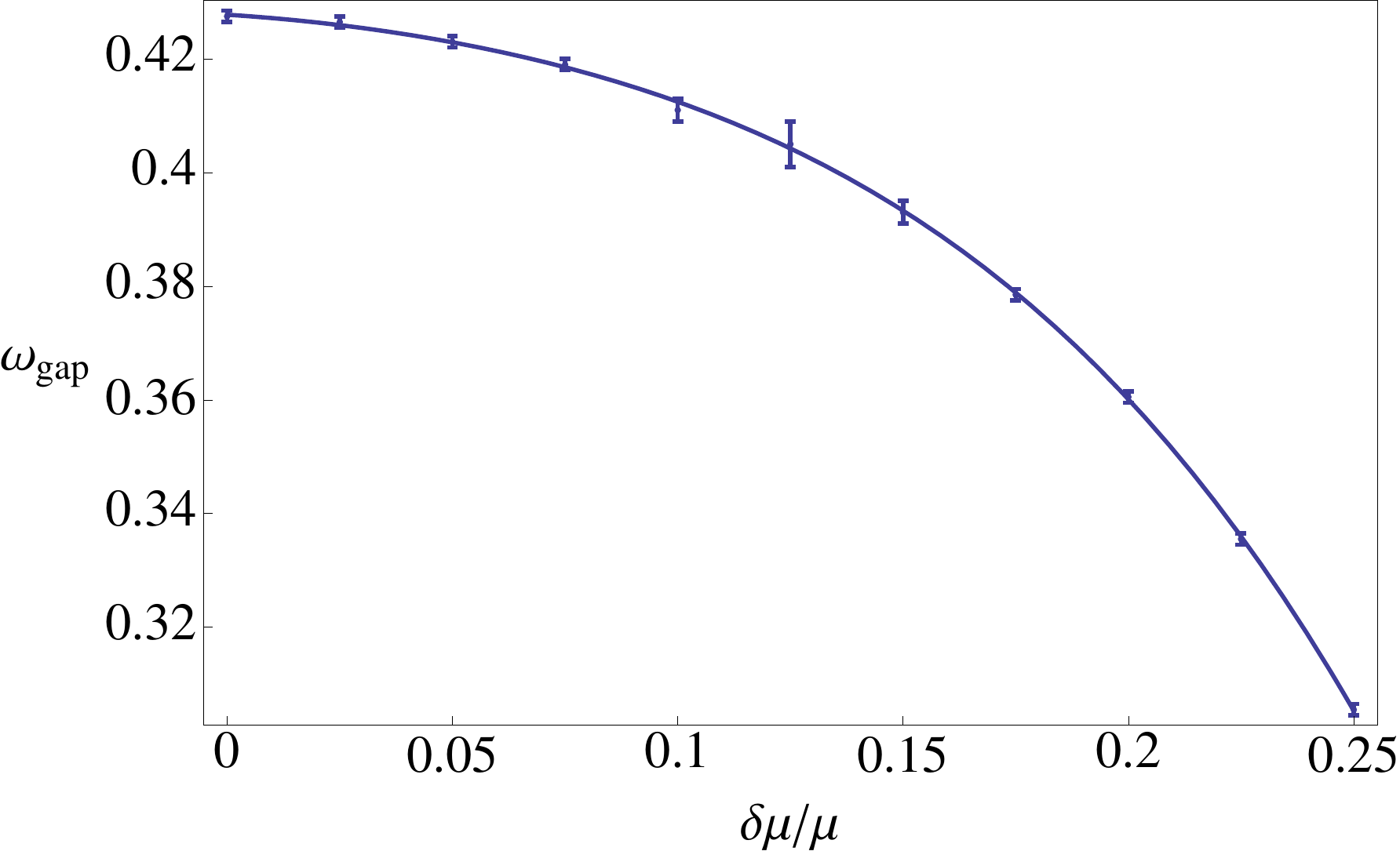}
\caption{\label{pseudogap} The ``pseudo-gap'' frequency $\omega_{\text{gap}}$ as a function of $\delta\mu/\mu$
at fixed $T$. The error bars are associated to numerical uncertainty and the plotted line emerges from a fit with a quartic polinomial.}
\end{figure}

Finally, in Figure \ref{pseudogap} we report the plot of the ``pseudo-gap'' frequency $\omega_{\text{gap}}$ below which the
real part of the optical electric conductivity $\sigma_A$ is essentially vanishing,\footnote{To be precise,
we used the numerical threshold value $\rm{Re}[\sigma_A(\omega_{\text{gap}})]=0.005$.} at constant temperature but increasing $\delta\mu/\mu$.
The behavior of $\omega_{\text{gap}}(\delta\mu/\mu)$ is clearly non-linearly decreasing.
This has to be compared to the case of ordinary unbalanced superconductors, where the gap $\Delta$ is constant in $\delta\mu$ at $T=0$.
In \cite{hhh2} it was pointed out that $\omega_{\text{gap}}(\delta\mu=0)/T_c^0 \sim 8$ as in some measures in high $T_c$ superconductors.
We find that $\omega_{\text{gap}}(\delta\mu/\mu)/T_c$ is not approximately constant, but a decreasing function of $\delta\mu/\mu$ which substantially deviates from the value 8 even in the range of $\delta\mu/\mu$ where it is still reasonably well defined.

\section{Comments on possible string embeddings}
\setcounter{equation}{0}
\label{embeddings}
If the results we have found on ``charge'' and ``spin'' transport properties have some degree of universality,
as we have argued above, the same is not necessarily true for equilibrium properties. The
$(T,\delta\mu)$ phase diagram, for example, can vary for different choices of the scalar potential, as
it happens in the balanced case (see for example \cite{diego}). This variation accounts for different
microscopic properties of the dual field theories which are however unknown. The latter could be
unvealed only by embedding the bottom-up models in full-fledged string or M-theory constructions.

The embeddings depend both on the spacetime dimensionality and on the microscopical details of the dual field theory - essentially on the kind of $U(1)$ gauge fields entering the game and on the precise nature of the order parameter. The latter can in fact be a condensate of, say, adjoint (or more generic two-index representations of some gauge group) or fundamental fermionic degrees of freedom and the string or M-theory embedding would strongly depend on whether one or another possibility is realized.
\subsection{Adjoint fermion condensates}
In supersymmetric contexts, this is a case where, say, gluino bilinears break some $U(1)_R$ symmetry of the theory.\footnote{$U(1)_R$ superfluidity driven by gluino Cooper pairs in ${\cal N}=4$ SYM at finite density has been first considered in \cite{michela}.} In this case we should try to see whether our minimal $3+1$ dimensional gravity model can arise from a consistent Kaluza-Klein truncation of 11d-supergravity on some compact seven-manifold, in the same way as it happens (see \cite{harden,gauntlett}) for the balanced model introduced in \cite{hhh1}. Isometries of the seven-manifold, in fact, are mapped into global (R-) symmetries of the dual field theory. At least within known consistent KK truncations (see e.g. \cite{gauntlett}) it seems that an embedding of a fully backreacted model containing the same fields as ours is possible only provided at least another non-trivial real scalar field is present. This is true also in the normal phase, i.e. in the absence of our complex scalar field.

To see this explicitly, let us consider the KK reduction of 11d-supergravity on a  seven sphere $S^7$. This gives rise to gauged $d=4$, ${\cal N}=8$ $SO(8)$ supergravity, which can be further truncated to a gauged ${\cal N}=2$ model where the bosonic sector consists of the metric, four Maxwell fields $A_{\mu}^A$, three ```dilatons'' $\phi_i$ and three ``axions'' \cite{cvetic}. The resulting Lagrangian density reads
\be
(\sqrt{\det g})^{-1}{\cal L} = R -\frac{1}{2}\sum_{i=1}^3\left[(\partial\phi_i)^2 + 8g^2(\cosh\phi_i)\right]-\frac{1}{4}\sum_{A=1}^4 \sum_{i=1}^3 e^{{a^i}_A\phi_i}(F^A)^2 + \dots\,,
\ee
where $g$ is the coupling, $a^i_A$ are constants and we have not included the contribution of the axions.
This action cannot be further truncated to a model with constant dilatons and two {\it independent} non-trivial Maxwell fields. Such kind of truncation is only allowed if the $U(1)$ fields are identified up to some constant (hence in the balanced case), and this constraint applies also to more general KK truncations \cite{gauntlett}.

The same conclusion holds in the $d=5$ case, i.e. considering abelian \cite{cvetic} or non-abelian gauged supergravities \cite{romans} obtained from consistent KK truncations of IIB supergravity.
For example in the first reference in \cite{gauntlett} it is shown how a $U(1)^2$-charged RN-$AdS$ black hole solution like the one describing the normal phase in our model, can be embedded within Romans' ${\cal N}=4$ $SU(2)\times U(1)$ gauged supergravity \cite{romans} (and thus in type IIB), only if the two Maxwell fields (and so the corresponding chemical potentials) are identified modulo a constant. Again, a setup where the two fields are independent can only be realized adding a non-trivial real scalar (dilaton-like) field. Embeddings of balanced superconductors \cite{IIBembe}, instead, do not require this extra scalar to be present.

\subsection{Fundamental fermion condensates}
\footnote{We are grateful to Emiliano Imeroni for his contributions to this part of the project.}
Matter fields transforming in the fundamental representation are introduced in the holographic correspondence by means of flavor D-branes. Models of holographic p-wave superconductivity have been actually embedded in probe flavor brane setups \cite{erme,erdmeng}. Here we want to focus on the s-wave case. Having in mind 4d QCD-like models, one could consider, say, a non-critical 5d string model with $N_c$ D3 and $N_f$ spacetime filling D4-anti-D4 branes \cite{paris,ckp}. The low-energy modes of the D3-branes would be the $SU(N_c)$ gluons, and the D4-anti-D4 branes would provide the left and right handed fundamental flavor fields (the quarks). The model contains a complex scalar field (the would-be tachyon of the open string stretching between branes and anti-branes) transforming in the (anti)fundamental of $SU(N_f)_L\times SU(N_f)_R$: its condensation (eventually triggered by a running dilaton which can account for confinement) drives the breaking of the chiral symmetry down to $SU(N_f)$ and thus it is dual to the chiral condensate of fundamental fermions \cite{sugimoto,paris,ckp}. The model (at least a simplified version of it \cite{paris}) can provide $AdS_5$ flavored solutions (at zero temperature and densities) with trivial tachyon and constant dilaton. These are possibly dual to phases of the theory in Banks-Zacks-like conformal windows. The corresponding finite temperature versions have been studied in \cite{unqqgp}.

Let us start considering the $N_f=1$ case. In addition to the above mentioned complex scalar field $\tau$,
the DBI brane-antibrane action \cite{sen,garousi} contains two Maxwell fields and it is coupled with the dilaton. The scalar field $\tau$ is charged under a combination of the two gauge fields (the ``chiral'' $U(1)_A$) and uncharged under the orthogonal combination (the ``baryonic'' $U(1)_B$). The chiral symmetry is actually anomalous and this is accounted for by other terms in the D-brane action \cite{ckp}. One can consider $N_f>1$ setups, e.g. $N_f=2$ ones, enhancing the preserved flavor symmetry to an $SU(2)\times U(1)_B$ one. Starting from the above mentioned flavored-$AdS$ solutions one could turn on a chemical potential for a $U(1)_I\subset SU(N_f)$ ``isospin'' field as well as a baryonic one. Analogously to our condensed matter setup, one could then study the stability of the system under fluctuations of the complex scalar field $\tau$. The latter would be dual to, say, a ${\bar u} d$ mesonic condensate and thus the role of $\delta\mu$ and $\mu$ would be played by $\mu_B$ and $\mu_I$ respectively. Let us focus on the $N_f=1$ case for simplicity, treating the axial $U(1)_A$ as if it were be a genuine symmetry replacing $U(1)_I$, i.e. neglecting all the terms in the brane action which are related to its anomaly.

The D$p$-brane-anti-brane DBI action in string frame reads \cite{sen,garousi}
\be
S = - T_p \int d^{p+1}x\ e^{-\Phi} V(|\tau|) \left[\sqrt{-\det{\mathbf A}^{(L)}}+\sqrt{-\det{\mathbf A}^{(R)}}\right]\,,
\label{senac}
\ee
where $T_p$ is the D-brane tension, $\Phi$ is the dilaton and $V(|\tau|)$ is the open string tachyon potential. Let us assume that a closed string tachyon, which could be present in non-critical Type 0 setups, is eventually frozen out. The matrices $\mathbf{A}^{(L),(R)}$ in (\ref{senac}) are defined as
\bea
&&{\mathbf A}^{(i)}_{MN}= P[g+B_{2}]_{MN} + 2\pi\alpha'F^{(i)}_{MN} + \pi\alpha'(D_M\tau)^*(D_N\tau)+ \pi\alpha'(D_N\tau)^*(D_M\tau)\,, \nonumber \\
&&F^{(i)}_{MN}=\partial_M A^{(i)}_N - \partial_N A^{(i)}_M\,,\qquad D_M\tau = (\partial_M + i A^{(L)}_M -i A^{(R)}_M)\tau\,,
\eea
where $i=L,R$ label the brane or antibrane, $P[\bullet]$ denotes the pullback on the D-brane worldvolume, $g$ is the metric and $B_{2}$ is the NSNS antisymmetric two-form. Notice that, as anticipated, the scalar field is charged only under the axial $U(1)_A$ combination
\be
A_M = A^{(L)}_M - A^{(R)}_M\,.
\label{axial}
\ee
It is instead uncharged under the barionic $U(1)_B$ combination
\be
B_M =A^{(L)}_M + A^{(R)}_M\,.
\label{vector}
\ee
In the case of brane-antibrane pairs in flat spacetime, string field theory gives a tachyon potential of the form
\be
V(|\tau|) = e^{\pi\alpha'm^2|\tau|^2}\,,\quad {\rm with}\,\,m^2=-\frac{1}{2\alpha'}\,.
\ee
This expression could be affected by non-trivial field redefinitions and by the fact that the branes have to be put on curved spacetimes (see e.g. a discussion in \cite{ckp}). Anyway, we will take this expression as a guideline.

As for the embedding described in the previous subsection, when independent Maxwell fields are both turned on, it seems not possible to have solutions with non-trivial dilaton (it is understood that the whole gravity action, say for $p=4$, will be a ``bulk+brane" one with the standard kinetic term for the dilaton). Notice that in the model the Maxwell fields are coupled by the non-linear structure of the DBI action and thus spintronics effects could be present also in the non-fully backreacted case.

Let us just notice, finally, that on a closed string background where $B_{2}=0$ and the metric is diagonal, the low energy effective Lagrangian density coming from (\ref{senac}) at the quadratic level in the fields is (see eq. (10) in the second paper in \cite{garousi} and use our redefinitions (\ref{axial}), (\ref{vector}))
\be
{\cal L}\approx -T_p (2\pi\alpha') e^{-\Phi} \sqrt{g}\left[-\pi\alpha'F^2 -\pi\alpha' Y^2 + |D\tau|^2 + m^2|\tau|^2\right]\,,
\ee
where $F=dA$, $Y=dB$ and $D\tau = (\partial -iA)\tau$. This has, modulo coefficients and the overall dilaton coupling, the same form as the matter part of our gravity model.

\vskip 15pt \centerline{\bf Acknowledgments} \vskip 10pt \noindent We are very grateful to Emiliano
Imeroni for his collaboration in the initial stages of this work and to Matteo Bertolini, Pasquale
Calabrese, Andrea Cappelli, Roberto Casalbuoni, Davide Forcella, Iroshi Kohno, Alberto Lerda, Massimo
Mannarelli, Alberto Mariotti, Mihail Mintchev, Giovanni Ummarino and Paola Verrucchi for many relevant
comments, advices, discussions and E-mail correspondence. This work was supported in part by the
MIUR-PRIN contract 2009-KHZKRX. The research of A. L. C. and F. B. is supported by the European
Community Seventh Framework Programme FP7/2007-2013, under grant agreements n. 253534 and 253937.

{ \it We would like to thank the Italian students,
parents, teachers and scientists for their activity in support of
public education and research.}

\appendix
\section{The Chandrasekhar-Clogston bound} \label{chandrase}
 Let us consider a Fermi mixture with two
species $u$ and $d$ (e.g. dressed electrons of spin up and down in metallic superconductors) with
different chemical potentials $\mu_u$ and $\mu_d$. Let us define the mean chemical potential $\mu$ and
the chemical potential imbalance $\delta\mu$ as \be \mu = \frac12 (\mu_u+\mu_d)\,\quad
\delta\mu=\frac12 (\mu_u-\mu_d)\,. \ee Assuming analiticity, in the grand-canonical ensemble, the
Gibbs free energy $\Omega(\delta\mu)$ at zero temperature can be Taylor expanded for $\delta\mu\ll\mu$
as
\begin{equation}
 \Omega(\delta\mu)=\Omega(0)+\Omega(0)'\delta\mu +\frac12 \Omega''(0)\delta\mu^2 + {\cal O}(\delta\mu^3)\,.
\label{expa}
\end{equation}
The first (resp. second) derivative of $\Omega$ w.r.t. $\delta\mu$ defines the population imbalance
$\delta n$ (resp. the susceptibility imbalance\footnote{In the case where the chemical potential
imbalance is induced by the Zeeman coupling with a magnetic field, this is actually the magnetic
susceptibility.} $\delta\chi$) \be \delta n\equiv n_u-n_d =
-\frac{\partial\Omega}{\partial\delta\mu}\,,\quad \delta\chi= \frac{\partial\delta
n}{\partial\delta\mu}=-\frac{\partial^2\Omega}{\partial\delta\mu^2}\,. \ee In BCS theory, the normal
phase at $T=0$ and $\delta\mu\ll\mu$ has a population imbalance given by $\delta n_N\approx
\rho_F\delta\mu$, where $\rho_F$ is the two-Fermion density of states at the mean Fermi surface
$E=E_F=\mu$. From this expression we easily find that, in the normal phase, the free energy expansion
(\ref{expa}) reduces at leading order to
\begin{equation}
\Omega_N(\delta\mu)\approx \Omega_N(0)-\frac12\rho_F\delta\mu^2\,.
\end{equation}
At $T=0$, the homogeneous BCS superconducting phase, characterized by Cooper pairs of zero total
momentum, has an equal number of particles of species $1$ and $2$: $\delta n_S=0$. Thus the free
energy just expands as
\begin{equation}
\Omega_S(\delta\mu)\approx\Omega_S(0).
\end{equation}
The difference between the two free energies is given by
\begin{equation}
\Omega_N(\delta\mu)-\Omega_S(\delta\mu)\approx \Omega_N(0)-\Omega_S(0)-\frac12\rho_F\delta\mu^2.
\end{equation}
Now, using the standard BCS result, $\Omega_N(0)-\Omega_S(0)=\rho_F\Delta_0^2/4$, where $\Delta_0$ is
the gap parameter at $T=0$, it follows that
\begin{equation}
\Omega_N(\delta\mu)-\Omega_S(\delta\mu)\approx\frac{1}{4}\rho_F\Delta_0^2-\frac12\rho_F\delta\mu^2.
\end{equation}
This shows that at $T=0$ the superconducting phase is favored (i.e. its free energy is less than the
free energy of the normal phase) only if the Chandrasekhar-Clogston bound
\begin{equation}
\delta\mu<\delta\mu_{1},\quad \delta\mu_{1}\equiv\frac{\Delta_0}{\sqrt{2}}\,,
\end{equation}
is satisfied.

\section{Equations of motion in  $d+1$ bulk spacetime
dimensions}\setcounter{equation}{0} \label{general} Let us consider
the generalization of our model to $d+1$-dimensions. The action
reads
\begin{equation}\label{gravlaggen}
S=\frac{1}{2k^{2}_{d+1}}\int dx^{d+1} \sqrt{-g}\left[ \mathcal{R}+ \frac{d(d-1)}{L^2}-
\frac{1}{4}F_{ab}F^{ab} - \frac{1}{4}Y_{ab}Y^{ab} - V(|\psi|)-|\partial \psi-iqA\psi|^{2}\right]\,.
\end{equation}
The ansatz for the spacetime metric is
\begin{equation}
 ds^{2}=-g(r)e^{-\chi(r)}dt^{2}+\frac{r^{2}}{L^{2}}d\vec{x}^2+\frac{dr^{2}}{g(r)}\,,
\end{equation}
 \noindent  together with an homogeneous ansatz for the fields
\begin{equation}\label{homogeneousgen}
\psi=\psi(r)\,, \quad A_a dx^a=\phi(r)dt\,, \quad B_a dx^a=v(r)dt\,.
\end{equation}
The equation of motion  for the scalar field reads
 \begin{equation}\label{psigen}
 \psi^{ \prime\prime}\!+\!\psi^\prime\Bigl(\frac{g\prime}{g}\!+\!\frac{(d-1)}{r}\!-\!\frac{\chi^\prime}{2}\Bigr)
\!-\!\frac{1}{2}\frac{V^\prime(\psi)}{g}\!+\!\frac{e^{\chi}q^{2}\phi^{2}\psi}{g^{2}}\!=\!0\,.
\end{equation}
Maxwell's equation for the $\phi$ field gives
\begin{equation}
 \phi^{\prime\prime}\!+\!\phi^\prime\Bigl(\frac{(d-1)}{r}\!+\!\frac{\chi^\prime}{2}\Bigr)\!-\!2\frac{q^{2}\phi\psi^{2}}{g}\!=\!0\,.
\end{equation}
Einstein's equations reduce to
\begin{eqnarray}
 &&\frac{1}{2}\psi^{\prime2} + \frac{e^{\chi}(\phi^{\prime2} + v^{\prime2})}{4g} + \frac{(d-1)}{2}\frac{g^\prime}{gr}+ \frac{1}{2}\frac{(d - 1)(d-2)}{r^{2}} - \frac{d(d-1)}{2gL^{2}} + \frac{V(\psi)}{2g} + \frac{q^{2}\psi^{2}\phi^{2}e^{\chi}}{2g^{2}} = 0\,,\nonumber \\
 &&\chi^\prime + \frac{2}{(d-1)}r\psi^{\prime2} + \frac{2}{(d-1)}r\frac{q^{2}\phi^{2}\psi^{2}e^{\chi}}{g^{2}} =0\,.
\end{eqnarray}
Finally, Maxwell's equations for the additional gauge field read
\begin{equation}
 v^{\prime\prime}\! +\!v^\prime\bigg(\frac{(d-1)}{r}\!+\!\frac{\chi^\prime}{2}\bigg)\!=\!0\,.
\end{equation}
\subsection{The normal phase}
The gravity solution corresponding to the normal phase in the dual $d$-dimensional
field theory is the $U(1)^2$ - charged Reissner-N\"{o}rdstrom (RN)-$AdS_{d+1}$ black hole
\begin{eqnarray}
ds^{2}&=&-g(r)dt^{2}+\frac{r^{2}}{L^{2}}d\vec{x}^{2}+\frac{dr^{2}}{g(r)},\label{Fcp3}\\
g(r)&=&\frac{r^{2}}{L^{2}}\bigg(1-\frac{r_{H}^{d}}{r^{d}}\bigg)+
\frac{1}{2}\frac{(d-2)}{(d-1)}(\mu^{2}+\delta\mu^2)\bigg(\frac{r_H}{r}\bigg)^{2(d-2)}\bigg(1-\bigg(\frac{r}{r_{H}}\bigg)^{d-2}\bigg),\label {Fcp4}\\
A_{t}&=&\mu\bigg( 1-\bigg(\frac{r_{H}}{r}\bigg)^{d-2}\bigg), \label{Fcp5}\\
B_t&=&\delta\mu\bigg( 1-\bigg(\frac{r_{H}}{r}\bigg)^{d-2}\bigg),
\end{eqnarray}
where $r=r_H$ is the position of the outer horizon.

The charge densities of the dual field theory are related to the subleading behavior of
the bulk Maxwell fields as
\begin{equation}\label{charges}
\rho=\frac{1}{2k_{d+1}^2}\frac{(d-2)\mu r_H^{d-2}}{L^{d-1}}\,,\qquad
\delta\rho=\frac{1}{2k_{d+1}^2}\frac{(d-2)\delta\mu r_H^{d-2}}{L^{d-1}}\,.
\end{equation}
The black hole temperature is
\begin{equation}\label{Fcp6}
T=\frac{r_H}{4\pi L^2}\left[d-\frac{(d-2)^2}{(d-1)}\frac{(\mu^2+\delta\mu^2)L^2}{2 r_H^2}\right]\,.
\end{equation}

The Gibbs free energy density (hence the pressure) is given by
\begin{equation}\label{rnentropy}
\omega= -p = -\frac{1}{2k_{d+1}^2}\frac{r_H^d}{L^{d+1}}\bigg( 1+\frac{(d-2)}{2(d-1)}\frac{(\mu^2+\delta\mu^2)L^2}{r_H^2}\bigg).
\end{equation}
Consistently, the charge densities in (\ref{charges}) are obtained as
\begin{equation}
\rho=-\frac{\partial \omega}{\partial \mu}\,,\qquad \delta\rho=-\frac{\partial \omega}{\partial \delta\mu}\,.
\end{equation}
The energy density is given by
\begin{equation}
\epsilon= \frac{d-1}{k_{d+1}^2}\frac{r_H^d}{L^{d+1}}\bigg( 1+\frac{(d-2)}{2(d-1)}\frac{(\mu^2+\delta\mu^2)L^2}{r_H^2}
\bigg)\,.
\end{equation}
This satisfies the relation $\epsilon=(d-1)p$ related to the
vanishing of the trace of the stress energy tensor. 

\subsubsection{Near horizon geometry}\label{nearhorizongeometry}

The Reissner-Nordstrom geometry is interesting as $T\rightarrow0$.
In this limit the horizon radius has a fixed value at
\begin{equation}
 r_H^2=\frac{1}{2d}\frac{(d-2)^2L^2\mu^2}{(d-1)}\, .
\end{equation}
To find the near horizon metric take the series Taylor expansion of the blackening factor
\begin{equation}
g(r)\simeq g(r_H)+g^\prime(r_H)\tilde{r}+\frac{1}{2}g^{\prime\prime}(r_H)\tilde{r}^2\,,
\end{equation}
where again $r=r_H+\tilde{r}$ with $\tilde{r}\rightarrow0$.
We find that
\begin{equation}\label{pappetta}
g(r_H)=0\,, \qquad g^\prime(r_H)\sim T=0\,,\qquad g^{\prime\prime}(r_H)=\frac{2d(d-1)}{L^2}\, .
\end{equation}
The near horizon metric reads then
\begin{equation}
ds^2_{\mbox{\scriptsize{near horizon}}}\simeq -d(d-1)\frac{\tilde{r}^2}{L^2}dt^2+\frac{r_H^2}{L^2}d\vec{x}^2+\frac{L^2}{d(d-1)\tilde{r}^2}d\tilde{r}^2\,,
\end{equation}
from which we  recognize the $AdS_2\times R^{d-1}$ metric. The $AdS_2$ radius squared is $L_{(2)}^2=L^2/(d(d-1))$.
\subsection{Criterion for instability}
Let us consider the stability of the above solution at $T=0$ under
fluctuations of the charged scalar field. The equation one has to
consider is given in (\ref{psigen}). The background is the extremal
doubly charged RN-$AdS_{d+1}$. In the near horizon limit it is easy to
show that the equation of motion for $\psi$ reduces to that of a
scalar field of effective mass
\be
m_{(2)}^2 = m^2 - \frac{2q^2}{1+x^2}\,,\quad x\equiv\delta\mu/\mu\,,
\ee
on an $AdS_2$ background of radius $L_{(2)}^2=L^2/(d(d-1))$.

The background is unstable in the limit if the BF bound in $AdS_2$
is violated, i.e. if $L_{(2)}^2m_{(2)}^2<-1/4$. This is equivalent
to the condition
\be
\left(1+\frac{\delta\mu^2}{\mu^2}\right)\left(m^2 + \frac{d(d-1)}{4}\right) < 2q^2\,,
\ee
which generalizes our formula (\ref{semi}) valid for $d=3$.

\end{document}